\documentclass[sn-nature,iicol]{sn-jnl}
\usepackage{graphicx}%
\usepackage{multirow}%
\usepackage{amsmath,amssymb,amsfonts}%
\usepackage{amsthm}%
\usepackage{mathrsfs}%
\usepackage[title]{appendix}%
\usepackage{xcolor}%
\usepackage{textcomp}%
\usepackage{manyfoot}%
\usepackage{booktabs}%
\usepackage{algorithm}%
\usepackage{algorithmicx}%
\usepackage{algpseudocode}%
\usepackage{listings}%
\usepackage{pifont,xcolor} 
\usepackage{pdfpages}
\usepackage{pgffor}
\theoremstyle{thmstyleone}%
\theoremstyle{thmstyletwo}%
\theoremstyle{thmstylethree}%
\usepackage[dvipsnames]{xcolor}
\usepackage{subfiles}
\usepackage{tikz}
\usepackage{pgfplots}
\usepgfplotslibrary{colorbrewer}
\pgfplotsset{compat=1.18}
\usetikzlibrary{positioning, calc, backgrounds, shapes.geometric, shapes.arrows, arrows.meta, shapes.symbols, decorations.text, fit}
\usetikzlibrary{external}
\usepgfplotslibrary{colormaps}
\usepgfplotslibrary{statistics}
\usepackage[most]{tcolorbox}
\usepackage{hyperref}
\begin{document}
\newcommand{\bigmac}{Big\,Mac\textsuperscript{\textregistered}}
\newcommand{\class}{{\textsf{Classic Burger\,}}}
\newcommand{\dbone}{{\textsf{Delicious Burger \nolinebreak[4]1\,}}}
\newcommand{\dbtwo}{{\textsf{Delicious Burger \nolinebreak[4]2\,}}}
\newcommand{\sbone}{{\textsf{Sustainable Burger \nolinebreak[4]1\,}}}
\newcommand{\sbtwo}{{\textsf{Sustainable Burger \nolinebreak[4]2\,}}}
\newcommand{\nutri}{{\textsf{Nutritious Burger\,}}}
\title[Generative Artificial Intelligence for burgers]
{{\sffamily{\bfseries{Generative Artificial Intelligence creates 
delicious, sustainable, and nutritious burgers}}}}
\author*[1]{\fnm{Vahidullah} \sur{Tac}}\email{vtac@stanford.edu}
\author[2]{\fnm{Christopher} \sur{Gardner}}\email{cgardner@stanford.edu}
\author[1]{\fnm{Ellen} \sur{Kuhl}}\email{ekuhl@stanford.edu}
\affil[1]{\orgdiv{Department of Mechanical Engineering}, \orgname{Stanford University}, \orgaddress{\city{Stanford}, \country{USA}}}
\affil[2]{\orgdiv{Prevention Research Center}, \orgname{Stanford University School of Medicine}, \orgaddress{\city{Stanford}, \country{USA}}}
\abstract{Food choices shape both human and planetary health; yet, designing foods that are delicious, nutritious, and sustainable remains challenging. Here we show that generative artificial intelligence can learn the structure of the human palate directly from large-scale, human-generated recipe data to create novel foods within a structured design space. Using burgers as a model system, the generative AI rediscovers the classic Big Mac without explicit supervision and generates novel burgers optimized for deliciousness, sustainability, or nutrition. Compared to the Big Mac, its delicious burgers score the same or better in overall liking, flavor, and texture in a blinded sensory evaluation conducted in a restaurant setting with 101 participants; its mushroom burger achieves an environmental impact score more than an order of magnitude lower; and its bean burger attains nearly twice the nutritional score. Together, these results establish generative AI as a quantitative framework for learning human taste and navigating complex trade-offs in principled food design.}
\maketitle
Food choices rank among the most consequential decisions humans make, with far-reaching implications for both personal and planetary health \cite{editorial24}. The global food system contributes substantially to climate change \cite{vermeulenClimateChangeFood2012}, drives land-use change and biodiversity loss \cite{foleyGlobalConsequencesLand2005}, depletes freshwater resources \cite{wadaGlobalDepletionGroundwater2010}, and pollutes terrestrial and aquatic ecosystems \cite{xu21}. Sustaining the Earth within a safe operating space for humanity requires a rapid transition toward more sustainable food systems \cite{springmannOptionsKeepingFood2018}.

The same system exerts a dominant influence on human health \cite{herforth25}. More than one billion people consume diets that fail to meet basic nutritional needs \cite{cantonhelenFoodAgricultureOrganization2021}, while many more experience micronutrient deficiencies \cite{he24}. Poor dietary patterns contribute substantially to the global burden of chronic non-communicable diseases \cite{popkinNutritionTransitionLowIncome2009}, 
including 
type II diabetes \cite{auneMeatConsumptionRisk2009}
cardiovascular disease \cite{lopezbarrera23} and 
certain cancers \cite{huangCardiovascularDiseaseMortality2012}. 
Diet thus links environmental sustainability and human health through shared consumption patterns \cite{lappe71}.

Consumer acceptance, rather than availability, remains the central bottleneck in the adoption of sustainable and nutritious foods \cite{friedrich22}. Deficits in 
taste \cite{laureati24}, 
texture \cite{st.pierreMimickingMechanicsComparison2024a}, and 
cultural familiarity \cite{mellorConsumerKnowledgeAcceptance2022}
continue to limit uptake, 
even when environmental and nutritional benefits are clear. Designing foods that satisfy environmental and nutritional objectives, while meeting sensory expectations, demands a quantitative understanding of the human palate \cite{spence15}. Yet, conventional food development relies heavily on artisanal expertise and incremental trial-and-error, which constrains systematic exploration of large design spaces and slows innovation \cite{kuhl25}.

To address this gap, we leverage generative artificial intelligence to model human food preferences as a high-dimensional probability distribution, with burgers as a model system. Rather than encoding rules for flavor or texture, the model learns statistical regularities directly from observed ingredient combinations and quantities. This probabilistic formulation enables discovery of novel recipes and optimization with respect to external criteria, including environmental impact and nutritional quality, while remaining grounded in consumer-relevant designs. These capabilities require that the model internalizes human taste preferences beyond superficial correlations.\\[6.pt]
\noindent{\it{Here we test the hypothesis that generative artificial intelligence can learn the human palate and create burgers that taste the same or better than the classic \bigmac.}} \\[6.pt]
The \bigmac ranks among the most widely consumed burgers worldwide \cite{spencer05}, 
it is sold in more than 100 countries \cite{McDonaldsDeliversStrong2011}, and 
its price serves as an informal index of purchasing power parity across currencies \cite{BigMacCurrencies1998}. 
This global familiarity makes the \bigmac a stringent benchmark to evaluate whether a generative model captures collective human taste preferences. To test this premise, we evaluate the model using two discovery benchmarks: the rediscovery of the \bigmac from statistical structure alone and the creation of novel burgers that retain high consumer acceptance.

Beyond validation, we integrate environmental sustainability and nutritional quality as additional criteria to select from a large ensemble of generated recipes. We quantify environmental impact using land use, greenhouse gas emissions, eutrophication potential, and scarcity-weighted water use \cite{clarkEstimatingEnvironmentalImpacts2022}, and assess nutritional quality using established profiling frameworks, including the healthy eating index \cite{krebs-smithUpdateHealthyEating2018}. By sampling broadly and selecting recipes that jointly satisfy palatability, sustainability, and nutrition, we show that generative artificial intelligence can identify burgers that substantially reduce environmental impact and improve nutritional quality without abandoning cultural familiarity. Sensory validation with more than 100 participants confirms that the model correctly captures human taste preferences and creates burger recipes that match or exceed the sensory appeal of the classic \bigmac.
\section*{{\sffamily{\bfseries{Results}}}}\label{sec_results}
{\sffamily{\bfseries{Generative AI successfully creates burger recipes.}}}
\begin{figure*}
    \centering
    \includegraphics[width=\textwidth]{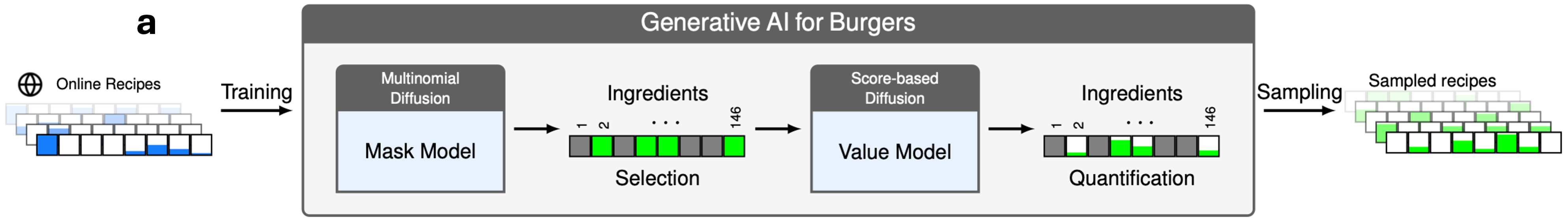} \\
    \includegraphics[width=\textwidth]{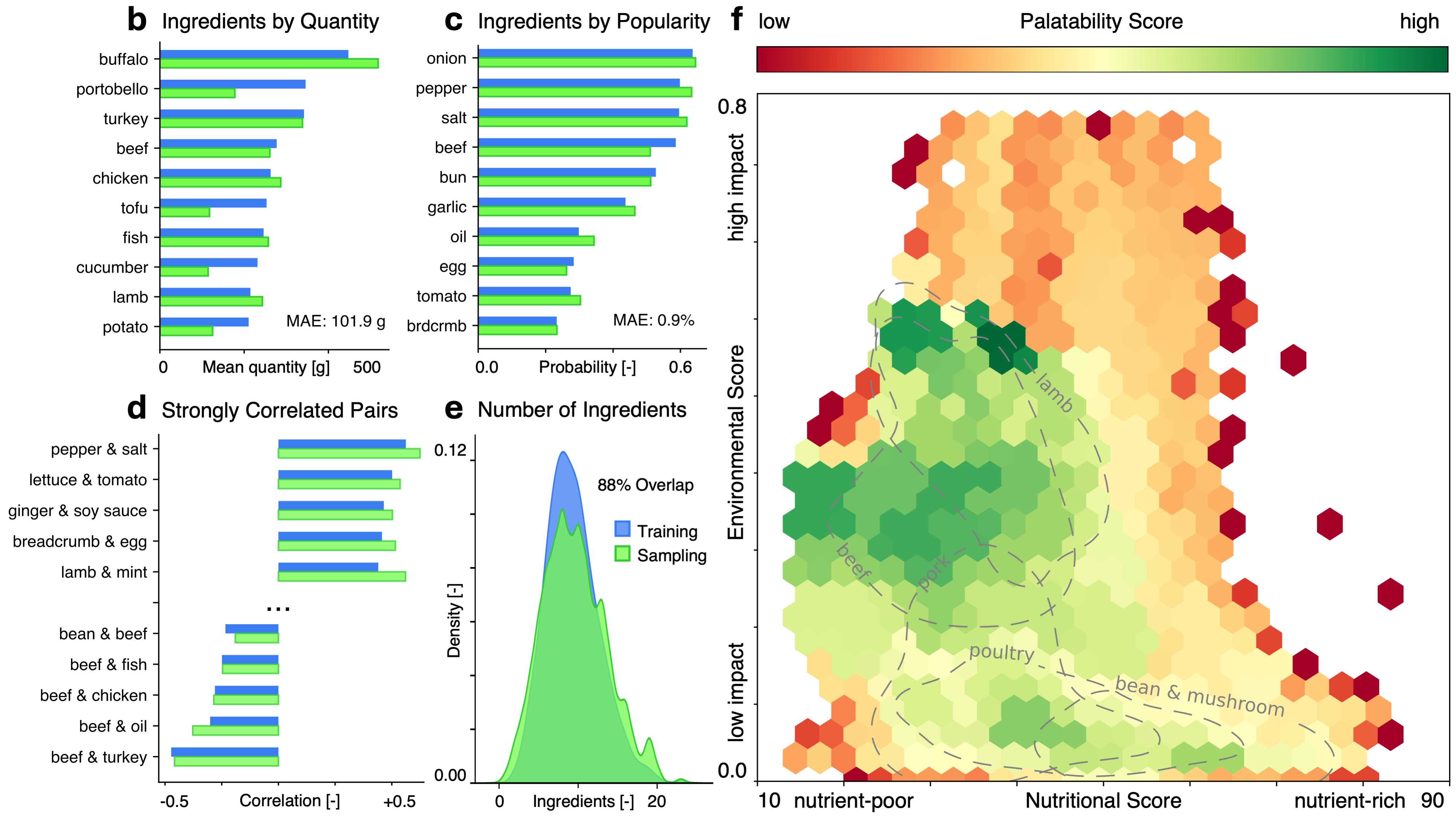}
    \caption{{\sffamily{\bfseries{Generative AI for Burgers.}}}
High-level model architecture with multinomial diffusion model for ingredient selection and score-based diffusion model for ingredient quantification 
(\textsf{a}); 
comparison of training and sampling data in terms of ingredients by quantity (\textsf{b}) and popularity (\textsf{c}); strongly correlated pairs (\textsf{d}); and total number of ingredients (\textsf{e}); 
distribution of palatability score, environmental score, and nutritional score in 1 million samples where a high palatability (green) indicates a popular ingredient combination and a low palatability (red) indicates a rare ingredient combination (\textsf{f}).}
    \label{fig1} 
\end{figure*}
We first validate the generative AI model by testing whether it reproduces key statistical properties of 2,216 human-designed burger recipes while generating novel combinations (Fig. \ref{fig1}). 
The model architecture combines a multinomial diffusion model for ingredient selection with a score-based generative model for ingredient quantification, which together generate complete burger recipes defined by 146 ingredients and their quantities (Fig. \ref{fig1}\textsf{a}). 
Comparison of generated samples with the training data shows close agreement across multiple marginal and higher-order statistics:
In particular, the model reproduces the distributions of ingredient quantities (Fig. \ref{fig1}\textsf{b}) and ingredient popularity, defined as the probability of ingredient occurrence across recipes (Fig. \ref{fig1}\textsf{c}), 
which demonstrates that it learns both 
how often ingredients appear and in what amounts. 
The model also captures higher-order structure, 
including strong positive and negative correlations between ingredient pairs commonly observed in real recipes (Fig. \ref{fig1}\textsf{d}), 
and accurately matches the distribution of recipe length, 
measured by the total number of ingredients per burger (Fig. \ref{fig1}\textsf{e}).
After establishing statistical fidelity, 
we generate one million burger recipes 
and map their palatability, environmental, and nutritional scores 
to reveal the structure of the generated design space (Fig. \ref{fig1}\textsf{f}). 
Recipes with high palatability scores 
cluster in regions associated with popular, conventional ingredient combinations
with lower nutritional and medium environmental scores,
whereas recipes with low palatability scores 
occupy regions associated with rare or unconventional combinations, 
consistent with human culinary preferences. 
Together, these results demonstrate 
that the generative model learns the underlying distribution 
of real human-designed burger recipes and 
enables systematic exploration 
of trade-offs between palatability, nutrition, and environmental impact.\\[6.pt]
{\sffamily{\bfseries{Generative AI rediscovers classic burgers and creates novel designs.}}}
\begin{figure*}[t]
    \centering
    \includegraphics[width=\textwidth]{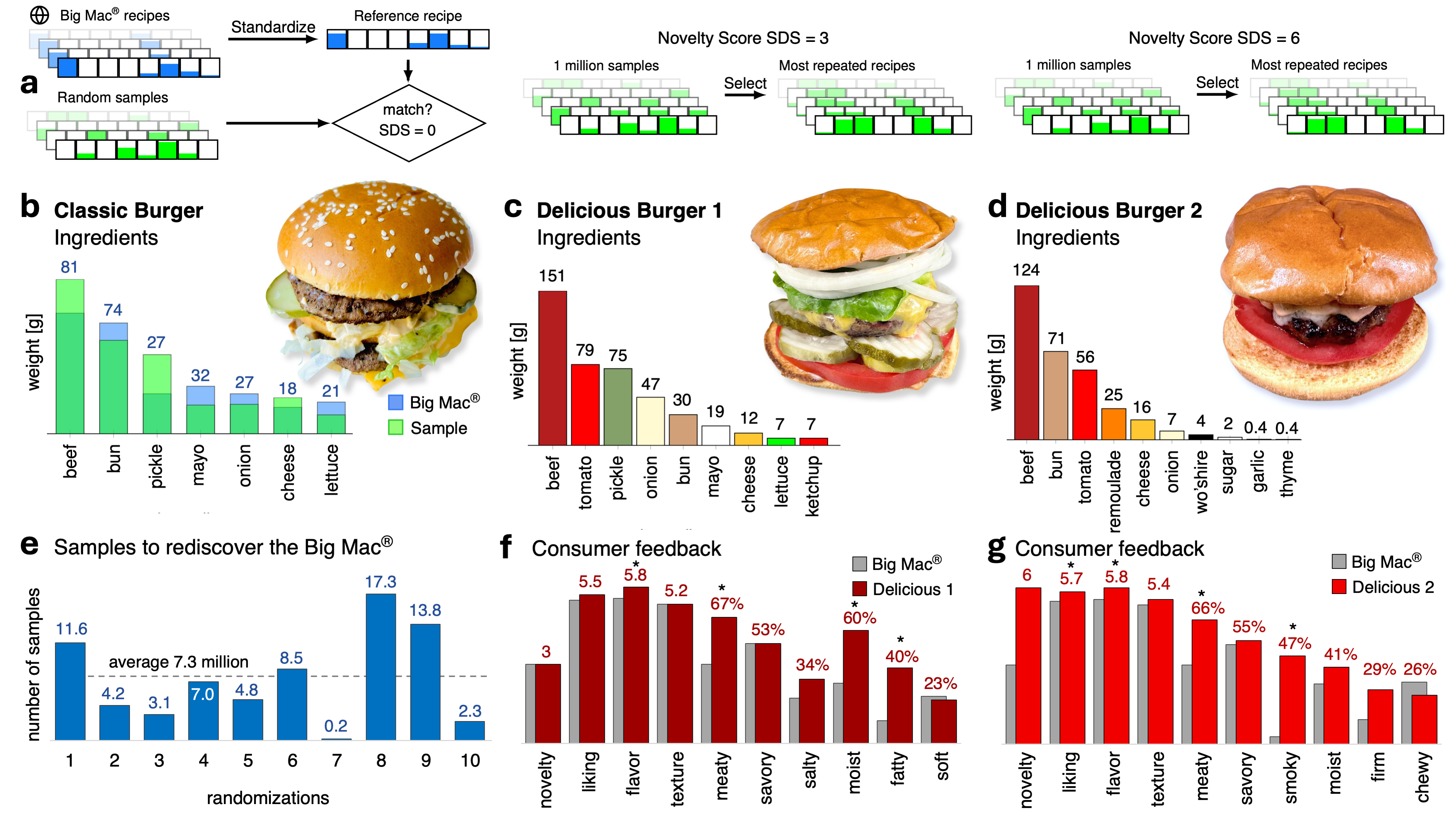}
    \caption{{\sffamily{\bfseries{Rediscovering Classic Burger and Discovering of Delicious Burgers.}}}
The generative AI 
rediscovers classic burgers and 
discovers novel burgers
from random AI generated samples, where 
a substantial difference score of zero, \textsf{SDS} = 0,
defines a match 
between a classic recipe and the sample, and
a substantial difference score larger than zero, 
\textsf{SDS} $>$ 0, defines the novelty score of the sample (\textsf{a});
rediscovered recipe of the \bigmac as representative \class (\textsf{b});
discovered recipes of \dbone with \textsf{SDS} $>$ 3 
and \dbtwo with \textsf{SDS} $>$ 6 
(\textsf{c},\textsf{d});
number of samples needed to rediscover the \bigmac for $n=$ 10 randomizations 
(\textsf{e}); 
consumer feedback for \dbone and \dbtwo compared to the \bigmac
(\textsf{f},\textsf{g});
$*$ indicates statistical significance ($p$\,$<$\,0.05).}
\label{fig2} 
\end{figure*}
We next assess whether the generative AI model can both rediscover canonical burger recipes and generate novel, appealing alternatives (Fig. \ref{fig2}). 
We quantify similarity between generated samples and a reference recipe 
using a substantial difference score, where 
\textsf{SDS} $=$ 0 indicates a match in ingredients and quantities, and 
\textsf{SDS} $>$ 0 measures increasing novelty 
(Fig. \ref{fig2}\textsf{a}). 
From random samples, 
the model successfully rediscovers the classic \bigmac,
both in correct ingredients and weights,
although the \bigmac was never part of the initial training data 
(Fig. \ref{fig2}\textsf{a}). 
Across ten independent randomizations, 
rediscovering the \bigmac requires on average 7.3 million samples
which demonstrates that exact replication of recipes
is a low-probability event 
under the learned distribution (Fig. \ref{fig2}\textsf{e}).
Beyond rediscovery, 
the model generates new burgers with varying degrees of novelty, 
illustrated by two representative recipes,
the \dbone with \textsf{SDS} $=$ 3 and 
the \dbtwo with \textsf{SDS} $=$ 6, 
which exhibit progressively more distinct ingredient profiles 
while retaining familiar burger structure 
(Figs. \ref{fig2}\textsf{c},\textsf{d}). 
Sensory evaluation indicates 
that the two delicious burgers achieve consumer ratings 
that are comparable to, and in some cases exceed, 
those of the classic 
\bigmac (Figs. \ref{fig2}\textsf{f},\textsf{g}):
\dbone received significantly higher ratings than the \bigmac
for flavor (5.8\,$\pm$\,1.3 vs 5.4\,$\pm$\,1.5) and
\dbtwo for
overall liking (5.7\,$\pm$\,1.2 vs 5.3\,$\pm$\,1.5) and 
flavor (5.8\,$\pm$\,1.3 vs 5.4\,$\pm$\,1.5), 
whereas both texture ratings 
did not differ significantly from the \bigmac 
($n$\,=\,101, 
$p$\,$<$\,0.05).
Participants more frequently described 
the \dbone 
as meaty  (67\% vs 42\%), moist (60\% vs 32\%), and fatty (40\% vs 12\%) 
than the \bigmac, and 
the \dbtwo 
as meaty (66\% vs 42\%) and smoky (47\% vs 4\%) 
than the \bigmac 
($n$\,=\,101; 
paired comparisons, 
$p$\,$<$\,0.05) 
Together, these examples validate the generative AI model 
by demonstrating its ability 
to internalize canonical burger recipes and 
generate novel, delicious, and palatable designs.\\[6.pt]
{\sffamily{\bfseries{Generative AI creates sustainable burgers.}}}
\begin{figure*}[t]
    \centering
    \includegraphics[width=\textwidth]{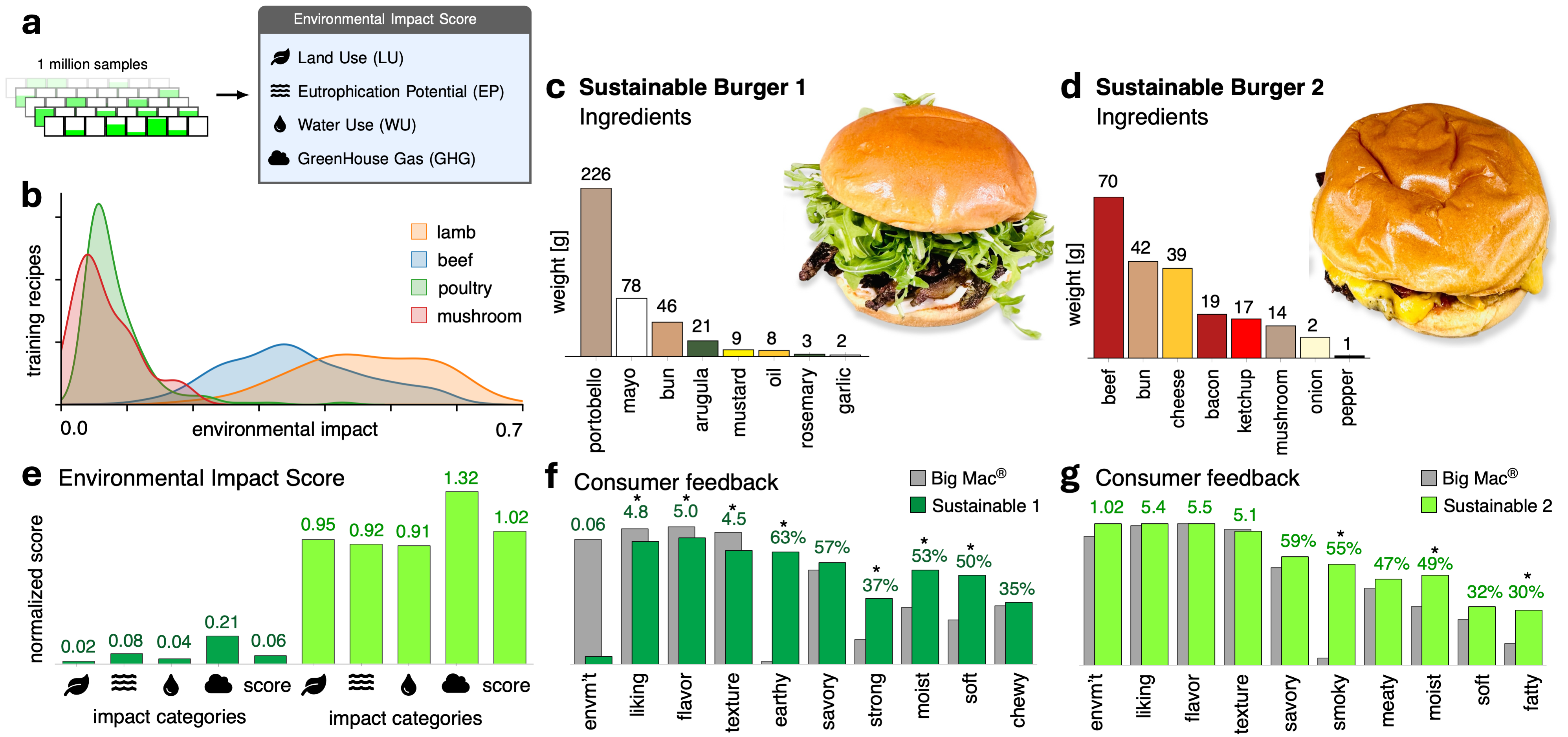}
    \caption{{\sffamily{\bfseries{Generating Sustainable Burgers.}}}
The generative AI 
generates sustainable burgers from
random AI generated samples, where 
the environmental impact score 
defines the collective impact 
of all ingredients in the recipe on
land use, 
eutrophication potential, 
scarcity-weighted water use, and 
greenhouse gas emissions (\textsf{a});
distribution of environmental impact score 
for all training recipes 
with lamb, beef, poultry, and mushroom
(\textsf{b});
generated recipes 
of \sbone and \sbtwo 
(\textsf{c},\textsf{d});
environmental impact scores of \sbone and \sbtwo
(\textsf{e});
consumer feedback for \sbone and \sbtwo 
compared to the \bigmac 
(\textsf{f},\textsf{g});
$*$ indicates statistical significance ($p$\,$<$\,0.05).}
    \label{fig3} 
\end{figure*}
We next evaluate 
whether the generative AI model 
can identify and generate burger recipes 
with reduced environmental impact 
while maintaining consumer acceptance (Fig. \ref{fig3}.) 
We quantify sustainability 
using an environmental impact score 
that aggregates ingredient-level contributions from 
land use, 
eutrophication potential, 
scarcity-weighted water use, and 
greenhouse gas emissions (Fig. \ref{fig3}\textsf{a}). 
Analysis of the training data shows 
that environmental impact scores vary substantially across recipes 
of different primary protein sources, 
with lamb- and beef-based recipes 
exhibiting systematically higher impacts 
than poultry- and mushroom-based recipes (Fig. \ref{fig3}\textsf{b}).
Sampling one million recipes from the model 
enables identification of sustainable burger candidates, 
illustrated by two representative examples, 
\sbone and \sbtwo, 
which differ in ingredient composition 
and dominant protein source (Fig. \ref{fig3}\textsf{c,d}). 
\sbone, a mushroom-based formulation, 
achieves an environmental impact score of 0.06,
more than one order of magnitude lower 
the \bigmac with 0.93, 
whereas 
\sbtwo, a mushroom-beef blend,
with 1.02 ranks comparable to the \bigmac
(Fig. \ref{fig3}\textsf{e}). 
Consumer feedback indicates 
that \sbone scores modestly below the \bigmac in overall liking, flavor, and texture, whereas \sbtwo performs on par with the \bigmac in these categories.
Sensory evaluation indicates 
that the two sustainable burgers achieve consumer ratings 
that differ in systematic ways from those of the classic \bigmac 
(Figs. \ref{fig3}\textsf{f},\textsf{g}):
\sbone received significantly lower ratings than the \bigmac for 
overall liking (4.8\,$\pm$\,1.8 vs 5.3\,$\pm$\,1.5), 
flavor (5.0\,$\pm$\,1.9 vs 5.4\,$\pm$\,1.5), and 
texture (4.5\,$\pm$\,1.9 vs 5.2\,$\pm$\,1.5), 
whereas ratings for 
\sbtwo did not differ significantly from the \bigmac 
across these attributes 
($n$\,=\,101; $p$\,$<$\,0.05).
Participants more frequently described 
the \sbone as 
earthy (63\% vs 2\%), strong (37\% vs 14\%), moist (53\% vs 32\%), and soft (50\% vs 25\%) than the \bigmac, and
the \sbtwo as 
smoky (55\% vs 4\%), moist (49\% vs 32\%), and fatty (30\% vs 12\%) 
($n$\,=\,101; 
paired comparisons, 
$p$\,$<$\,0.05 for all reported attributes).
Together, 
these results validate 
that the generative AI model 
successfully navigates the trade-off 
between sustainability and palatability
and discovers burgers 
with markedly reduced environmental impact 
without compromising taste.\\[6.pt]
{\sffamily{\bfseries{Generative AI creates nutritious and personalized burgers.}}}
\begin{figure*}[t]
    \centering
    \includegraphics[width=\textwidth]{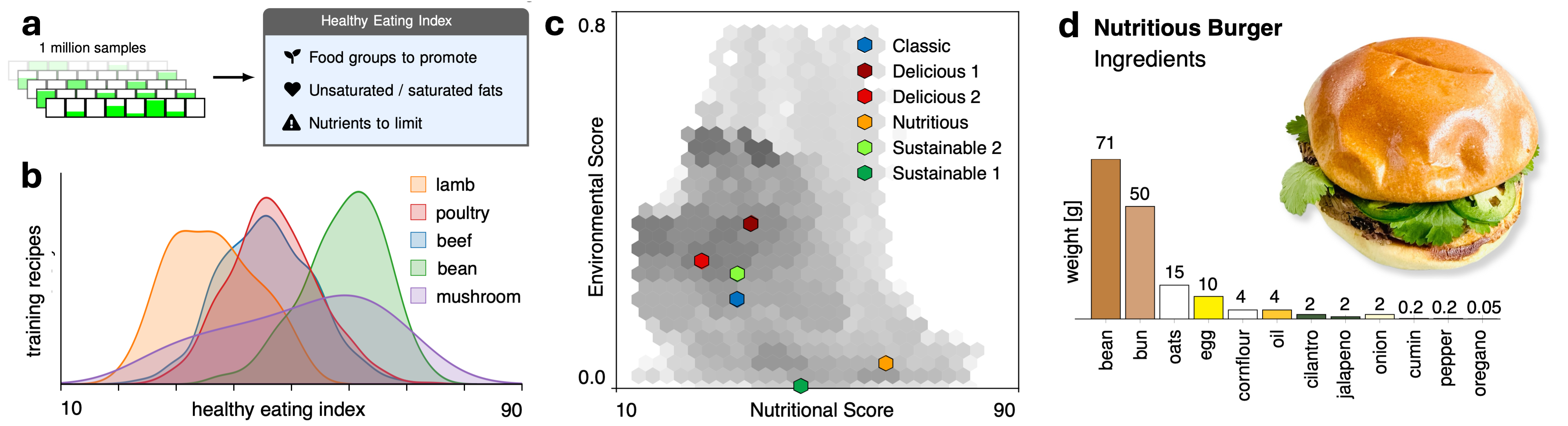}
    \includegraphics[width=\textwidth]{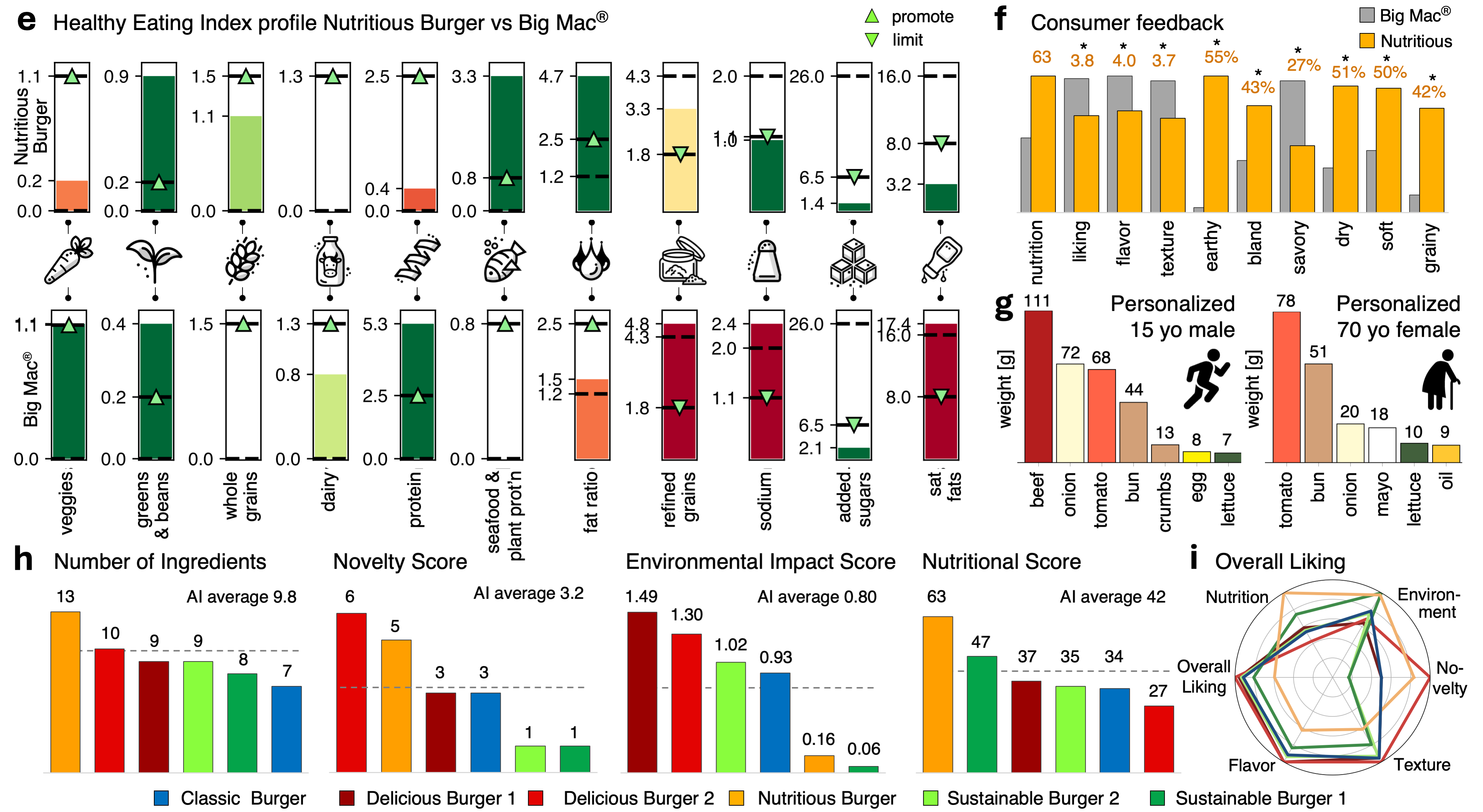}
    \caption{{\sffamily{\bfseries{Generating Nutritious and Personalized Burgers.}}}
The generative AI 
generates nutritious burgers from
random AI generated samples, where 
the healthy eating index 
defines the collective impact 
of all ingredients in the recipe on
food groups to promote, 
unsaturated and saturated fats, and 
nutrients to limit 
(\textsf{a});
distribution of healthy eating index
for all training recipes 
with lamb, poultry, beef, bean, and mushroom
(\textsf{b});
nutritional-environmental score of all six burgers
(\textsf{c});
generated recipe of \nutri 
(\textsf{d});
healthy eating index profile
of \nutri compared to the \bigmac
(\textsf{e});
consumer feedback for \nutri  
compared to the \bigmac 
(\textsf{f});
generated 
personalized recipes for 
15-year-old, 180\,cm, 80\,kg active male and 
70-year-old, 170\,cm, 70\,kg moderately active female 
(\textsf{g});
direct comparison of all six burgers by
number of ingredients, 
novelty, 
environmental impact, and
nutrition score
(\textsf{h});
overall comparison 
(\textsf{i});
$*$ indicates statistical significance ($p$\,$<$\,0.05).}
\label{fig4} 
\end{figure*}
We next evaluate whether the generative AI model 
can identify burger recipes optimized for nutritional quality 
and adapt them to individual dietary needs (Fig. \ref{fig4}). 
We quantify nutrition using the healthy eating index, 
which aggregates contributions from food groups to promote, fatty acid composition, and nutrients to limit (Fig. \ref{fig4}\textsf{a}). 
Analysis of the training data reveals substantial variation 
in healthy eating index 
across recipes with different primary protein sources, 
with bean- and mushroom-based recipes exhibiting systematically higher nutritional scores than beef- and lamb-based recipes (Fig. \ref{fig4}\textsf{b}). 
Sampling one million recipes enables 
identification of a representative \nutri 
with a high nutritional score, 
which occupies a favorable region 
of the nutritional–environmental design space 
compared to the other AI generated burgers (Fig. \ref{fig4}\textsf{c},\textsf{d}).
The \nutri, 
a bean-based formulation, 
achieves a healthy eating index of 63.12, 
nearly twice as high as the \bigmac with 33.71,
while also reducing its environmental impact score 
by a factor of six (Fig. \ref{fig4}\textsf{c}).
Relative to the \bigmac, 
the \nutri shows improved alignment with dietary guidelines 
across multiple healthy eating index components, 
including increased contributions from vegetables, whole grains, and plant protein, alongside reduced refined grains, sodium, and saturated fat (Fig. \ref{fig4}\textsf{e}).
Consumer feedback reveals a clear reduction 
in hedonic ratings for the \nutri 
relative to the \bigmac (Fig. \ref{fig4}\textsf{f}):
The \nutri received significantly lower ratings than the \bigmac for 
overall liking (3.8\,$\pm$\,1.7 vs 5.3\,$\pm$\,1.5), 
flavor (4.0\,$\pm$\,1.8 vs 5.4\,$\pm$\,1.5), and 
texture (3.7\,$\pm$\,1.8 vs 5.2\,$\pm$\,1.5)
($n$\,=\,101; $p$\,$<$\,0.05).
Participants more frequently described the \nutri as 
earthy (55\% vs 2\%), bland (43\% vs 21\%), dry (51\% vs 18\%), 
soft (50\% vs 25\%), and grainy (42\% vs 7\%), 
and less savory (27\% vs 53\%)
than the \bigmac 
($n$\,=\,101; paired comparisons, $p$\,$<$\,0.05 for all reported attributes).
Beyond population-level optimization, 
the AI model 
also generates personalized burger recipes 
tailored to individual nutritional requirements. 
We demonstrate this feature 
by producing personalized recipes for 
a highly active 15-year-old male and 
a moderately active 70-year-old female, 
which differ in ingredient composition and quantities 
in accordance with age- and activity-specific dietary needs (Fig. \ref{fig4}\textsf{g}). 
Finally, a direct comparison across all six burgers 
highlights systematic trade-offs between 
ingredient count, novelty, environmental impact, and nutrition, 
and places the AI generated burgers in distinct regions 
of this multi-objective design space (Fig. \ref{fig4}\textsf{h}). 
In this design space,
\dbone and \dbtwo
score best in overall liking, flavor, and texture,
whereas
\sbone and \nutri
score best in nutrition and environment
(Fig. \ref{fig4}\textsf{i}). 
Together, these results validate 
that the generative AI model 
can optimize for nutritional quality 
at both the population and individual levels 
while maintaining sensory acceptance.
\section*{{\sffamily{\bfseries{Discussion}}}}\label{discussion}
{\sffamily{\bfseries{Learning the structure of food design.}}}
A central result of this study is that generative AI can learn the latent structure of food design directly from large-scale, human-generated recipe data \cite{datta25}.
Rather than reproducing superficial statistics, the model captures higher-order regularities that define burgers as a culinary class, including ingredient co-occurrence, typical recipe length, and characteristic quantity distributions. 
This perspective echoes broader findings from data-driven culinary science, such as the flavor network approach that uncovers principled patterns of ingredient co-occurrence across tens of thousands of recipes in global cuisines \cite{ahn11}. 
The ability to internalize the {\it{statistical grammar of recipes}} reflects the high-dimensional and nonlinear nature of human taste, which emerges from complex interactions among ingredients rather than simple additive rules \cite{barabasi20}. 
Importantly, the rediscovery of classic benchmarks such as the \bigmac--without explicit supervision--demonstrates that culturally dominant foods occupy high-probability regions of the learned design space and validates the model against widely recognized culinary reference points. 
Together, these findings support the view that recipes encode collective human taste knowledge accumulated over decades, and that generative models can extract this knowledge to form an {\it{interpretable and navigable design manifold}}. 
This reframes food formulation from an artisanal, trial-and-error practice into a data-driven design science and positions recipes as a natural and human-centered interface between culinary tradition and artificial intelligence.\\[6.pt]
{\sffamily{\bfseries{Exploring novelty without sacrificing palatability.}}}
Beyond learning existing culinary structure, the generative model enables controlled exploration of novel burger designs while preserving sensory appeal. By explicitly quantifying novelty through a substantial difference score, we show that departures from canonical recipes do not inherently lead to diminished consumer acceptance. Instead, the model identifies regions of the design space where novelty and palatability coexist: It discovers burgers that differ meaningfully in ingredient composition and proportions while maintaining ratings for liking, flavor, and texture comparable to widely consumed references. This behavior reflects the {\it{high-dimensional and nonlinear nature of human taste}}, which emerges from complex interactions among ingredients and cannot be navigated reliably through intuition or simple empirical rules alone \cite{spence15}. It also aligns with systematic analysis showing that consumer acceptance of novel foods depends on intrinsic sensory properties such as taste and familiarity as well as psychological factors such as neophobia and food experience \cite{laureati24}. The rediscovery of classic burgers alongside successful novel variants suggests that cultural familiarity anchors broad regions of acceptability within the culinary design space that enable innovation without alienating consumer expectations \cite{torrico19}. In this context, generative AI provides a {\it{principled framework for exploring non-linear trade-offs}} between familiarity and novelty, and reframes culinary innovation as navigation of a structured, data-driven design manifold rather than trial-and-error experimentation.\\[6.pt]
{\sffamily{\bfseries{Environmental sustainability as a system-level design objective.}}}
A key contribution of this work is demonstrating that generative AI can substantially reduce the environmental impact of burgers while remaining grounded in consumer-relevant designs. Livestock production is a major driver of greenhouse gas emissions, land use, and biodiversity loss \cite{pooreReducingFoodsEnvironmental2018}; yet, consumer adoption of lower-impact alternatives remains limited by taste, texture, and familiarity \cite{godfray18}. Our results show that sustainability emerges not from isolated ingredient substitutions, but from coordinated changes across entire recipes. This yields designs that achieve large reductions in environmental impact through {\it{system-level rebalancing of ingredients}}. This finding aligns with broader evidence that dietary sustainability depends on integrated food-system redesign rather than single-ingredient replacements \cite{willett19}. Notably, the mushroom-only burger achieves an environmental impact score more than an order of magnitude lower than that of a conventional reference burger, while the mushroom-blended burger retains comparable impact and sensory performance. This highlights that environmental performance depends on interactions among ingredients, rather than any single component, and that generative models are well suited to navigating such non-linear, multi-objective trade-offs. By embedding life-cycle considerations directly into the generative process, the model identifies sustainable designs that remain recognizable as burgers and addresses a central bottleneck in alternative protein adoption--{\it{acceptance rather than availability}}--a challenge repeatedly emphasized in studies of consumer response to plant-based and hybrid meat alternatives \cite{bryant20}. Together, these findings bridge the gap between environmental metrics and consumer-facing design, and position generative AI as a practical tool for system-level optimization of food sustainability.\\[6.pt]
{\sffamily{\bfseries{Nutrition and personalization as explicit design criteria.}}}
An additional contribution of this work is demonstrating that nutritional quality can be treated as an explicit, quantitative design objective, rather than a secondary outcome of food formulation. By optimizing generated recipes with respect to the healthy eating index \cite{krebs-smithUpdateHealthyEating2018}, the model identifies burgers that substantially outperform widely consumed benchmarks, achieving nearly twice the nutritional score of a conventional reference burger, while simultaneously reducing environmental impact by a factor of six. This result underscores that improvements in nutrition and sustainability need not be mutually exclusive, but instead emerge from coordinated, system-level adjustments across ingredient composition and quantities \cite{willett19}. Importantly, the AI-generated nutritious burger achieves this performance without increasing reference complexity, which directly addresses growing {\it{concerns around ultra-processed foods}} and long ingredient lists that undermine consumer trust. Extending beyond population-level optimization, the AI model demonstrates the ability to translate dietary reference intakes into concrete, consumer-facing food designs by {\it{personalizing recipes based on age, sex, and activity level}}, building on earlier initiatives to personalize nutrition based on glucose levels \cite{zeevi15}. 
Together, these findings position generative AI as a practical bridge between nutritional science and everyday eating and enable objective comparison, optimization, and personalization of foods within a unified design framework.\\[6.pt]
{\sffamily{\bfseries{Limitations and scope.}}}
This study has several limitations that define the scope of the reported results. First, the generative model learns from existing, human-designed recipes and therefore inherits cultural, regional, and temporal biases present in the source data. Second, the recipe representation includes only ingredient identities and quantities and does not explicitly account for processing steps, cooking methods, or physical transformations that influence texture and flavor. Third, the environmental and nutritional scores rely on aggregated databases and global averages and therefore do not reflect variability associated with specific production practices or supply chains. Fourth, sensory validation focused on a limited set of generated burgers and participants, and broader studies will be required to establish generalizability across populations and contexts. Despite these constraints, the model provides a flexible and extensible framework for multi-objective food design.\\[6.pt]
{\sffamily{\bfseries{Broader implications and future outlook.}}}
This work demonstrates how generative AI can shift food formulation from artisanal trial-and-error toward a quantitative, data-driven design science. By learning directly from large-scale recipe data, the AI model functions as a domain-specific world model of burger design as it captures the statistical regularities and trade-offs that shape plausible and desirable foods. This enables systematic exploration of deliciousness, nutrition, and sustainability while remaining grounded in cultural familiarity, with adoption--rather than novelty--emerging as the central challenge in food innovation. More generally, food offers a uniquely human-centered domain for generative AI, where models can align optimization objectives with sensory feedback, health, and environmental constraints. Extending such approaches to incorporate processing, cooking, and supply-chain dynamics could ultimately support end-to-end food design and establish generative AI as a collaborative tool at the intersection of human creativity, engineering, and planetary health.
\section*{{\sffamily{\bfseries{Conclusion}}}}\label{conclusion}
This work shows that generative AI can reimagine food formulation as a quantitative, data-driven design process rooted in human preferences and measurable constraints. By learning directly from large-scale recipe data, the model captures the structure of burger design and enables systematic exploration across taste, nutrition, and sustainability. Consumer feedback validates the generative model by confirming that its designs align with human taste preferences while delivering substantial improvements in nutritional quality and environmental impact relative to conventional benchmarks. Beyond burgers, this approach points toward a new paradigm for AI-driven food design that unites culinary creativity with human and planetary health.

\small
\section*{{\sffamily{\bfseries{Methods}}}}\label{methods}
{\sffamily{\bfseries{\large{Model\,architecture,\,training\,and\,validation}}}} \\[6.pt]
Food recipes are hybrid discrete--continuous objects that require 
\emph{ingredient selection} and 
\emph{ingredient quantification}. 
Here we represent each burger recipe 
solely by its ingredients and their associated weights. 
Accordingly, a recipe 
$
\mathbf{x}_0 = \{ \mathbf{m}, \mathbf{w} \},
$
consists of a binary ingredient mask
$\mathbf{m}\in\{0,1\}^K$ 
that indicates presence or absence of each ingredient and 
the ingredient weight $\mathbf{w}\in\mathbb{R}^K$.
We adopt a two-stage diffusion-based framework 
that decouples ingredient selection from ingredient quantification:
Specifically, we integrate 
a \emph{multinomial diffusion model}\,\cite{NEURIPS2021_67d96d45} 
to generate ingredient masks  
with a \emph{score-based generative model}\,\cite{songScoreBasedGenerativeModeling2021} 
to generate ingredient weights 
conditional on a given mask (Fig.\,\ref{fig1}\textsf{a}).\\[6.pt]
{\sffamily{\bfseries{Diffusion-based recipe generation.}}}
Diffusion models 
define a forward stochastic process $q(\mathbf{x}_t|\mathbf{x}_{t-1})$ 
that gradually adds noise to data samples $\mathbf{x}_0$
to produce latent variables $\mathbf{x}_t$ 
that increasingly obscure the original data
throughout $t=1,...,T$ time steps \cite{NEURIPS2020_4c5bcfec}.
The learnable component 
is the reverse process $p(\mathbf{x}_{t-1}|\mathbf{x}_t)$, 
which progressively removes noise 
and enables generation from an unstructured prior.
We train the diffusion model
by maximizing a variational lower bound on the data likelihood,
which corresponds to the evidence lower bound,
\[
\begin{array}{l}
    \log \textsf{P}(\mathbf{x}_0) \\ 
    \quad \geq 
    \mathbb{E}_{\mathbf{x}_1,\cdots,\mathbf{x}_T \sim q} 
    \left[
        \log p(\mathbf{x}_T) 
        + \sum_{t=1}^T 
        \log \frac{p(\mathbf{x}_{t-1}|\mathbf{x}_t)}
        {q(\mathbf{x}_t|\mathbf{x}_{t-1})}
    \right]. 
\end{array}
\]
Here 
$\textsf{P}(\mathbf{x}_0)$
denotes the marginal likelihood of an observed burger recipe 
under the generative model, 
obtained by integrating over all latent diffusion variables, where
$\mathbf{x}_0$ is a human-designed burger recipe, 
$\mathbf{x}_t$ is its progressively noised representation, 
$q$ is the fixed forward process, 
$p$ is the learned reverse denoising process, and
$\mathbb{E}$ is the expectation operator that denotes an average over noise realizations drawn from the forward diffusion process.\\[6.pt]
{\sffamily{\bfseries{Ingredient selection via multinomial diffusion.}}}
We model ingredient selection using a multinomial diffusion process in which ingredient presence is treated as a categorical variable. We define the forward process as
\[
\label{eq_categorical}
    q(\mathbf{x}_t|\mathbf{x}_{t-1}) 
    = \mathcal{C}\!\left(
    \mathbf{x}_t \,\middle|\,
    (1-\beta_t)\mathbf{x}_{t-1} + \beta_t/K
    \right),
\]
where 
$\mathcal{C}$ denotes a categorical distribution
with the parameter listed after the vertical bar $|$, 
$\beta_t$ controls the noise level at time step $t$, and 
$K$ is the number of categories.
In our application, ingredient selection is \emph{binary}, $K=2$, meaning an ingredient is either present or absent. The above equation reduces to a Bernoulli distribution with parameter
$(1-\beta_t)\mathbf{x}_{t-1} + \beta_t(1-\mathbf{x}_{t-1})$,
which flips ingredient inclusion from present to absent or vice versa with a probability $\beta_t$ and keeps it the same with a probability $(1-\beta_t)$. 
As $t$ increases, the ingredient mask becomes progressively randomized, 
while the learned reverse process 
reconstructs statistically plausible ingredient combinations, 
which inherently capture dependencies 
between ingredients that commonly co-occur in burger recipes.\\[6.pt]
{\sffamily{\bfseries{Ingredient\,quantification\,via\,score-based\,diffusion.}}}
Conditional on a given ingredient mask, we generate ingredient quantities using a score-based generative model formulated through stochastic differential equations. 
The forward {\it{noising process}} is
\[
    \mathrm{d}\mathbf{x}_t 
    = f(\mathbf{x}_t,t)\,\mathrm{d}t 
    + g(t)\,\mathrm{d}B_t,
\]
and the reverse-time {\it{denoising process}} is
\[
    \mathrm{d}\mathbf{x}_t 
    = \left[
        g^2(t)\nabla_{\mathbf{x}}\log p_t(\mathbf{x})
        - f(\mathbf{x}_t,t)
    \right]\mathrm{d}t 
    + g(t)\,\mathrm{d}\tilde{B}_t,
\]
where 
$B_t$ is a $K$-dimensional Brownian motion, 
$\tilde{B}_t$ is its time reversal, and 
$f$ and $g$ define the drift and diffusion coefficients\,\cite{songScoreBasedGenerativeModeling2021, tacGenerativeHyperelasticityPhysicsinformed2024}.
Here,
rather than learning the probability density $p_t(\mathbf{x})$ directly,
we approximate the score function $\nabla_{\mathbf{x}}\log p_t(\mathbf{x})$ 
using a neural network 
to enable efficient sampling of ingredient weights 
consistent with observed distributions in human-designed burger recipes.
\\[6.pt]
{\sffamily{\bfseries{Dataset and training.}}}
We train our model on a curated burger dataset derived from an open-source collection of over half a million human-designed recipes from \texttt{Food.com}\,\cite{alvinFoodcomRecipesReviews2020, weialexanderFoodcomRecipesIngredients2023}. 
We filter all recipes for burgers,
extract ingredients, quantities, and units from free texts, and
standardize and convert the data
into a structured representation. 
The final dataset consists of 
2,216 burger recipes
made up of 
146 ingredients. 
Supplementary File\,1 provides additional details about the preprocessing step.\\[6.pt]
{\sffamily{\bfseries{Model validation and statistical fidelity.}}}
The trained model accurately reproduces both first-order and higher-order statistical properties of the training data:
The {\it{ingredient selection model}} 
estimates the marginal probability 
of each ingredient appearing in a random burger recipe 
with a maximum absolute error below 1\% (Fig.\,\ref{fig1}\textsf{b}).
The {\it{ingredient quantification model}} 
predicts quantities in previously unseen recipes 
with a mean absolute error 
of 101.9\,g (Fig.\,\ref{fig1}\textsf{c}),
despite the extrapolatory and highly stochastic nature of the problem.
Beyond marginal statistics, 
the model captures higher-order structure, 
including 
{\it{pairwise ingredient correlations}} (Fig.\,\ref{fig1}\textsf{d}) 
and 
{\it{number of ingredients per recipe}} (Fig.\,\ref{fig1}\textsf{b},\textsf{e}). 
These properties are not explicitly enforced during training, 
but emerge from the learned generative process. 
Supplementary File\,1 provides additional validation results and benchmarks.\\[6.pt]
{\sffamily{\bfseries{Substantial\,Difference\,Score\,to\,quantify similarity.}}}
Quantifying the proximity between recipes 
and grouping similar recipes 
is useful for various applications,
for example, to quantify the novelty of an AI-generated recipe. 
For this purpose,
we define the semi-discrete \emph{substantial difference score} 
between two recipes $r_1$ and $r_2$, 
\[
    \textsf{SDS} (r_1, r_2) = \sum_{i=1}^{n_\mathrm{ing}} d_i (r_{1}, r_{2}),
\] 
as the sum of the binary distance $d_i$ over all ${n_\mathrm{ing}}=146$ ingredients in the database, with
\[
  d_i (r_{1}, r_{2}) = \begin{cases}
    1 & \text{if } r_{1i} + r_{2i} \neq 0 \text{ and } r_{1i} \cdot r_{2i} = 0 \\
    1 & \text{if } \max(r_{1i}, r_{2i})/\min(r_{1i}, r_{2i}) \geq 2 \\
    0 & \text{otherwise}.
    \end{cases}
\]
During {\it{rediscovery}},
we use the substantial difference score of zero, $\textsf{SDS}=0$, 
to quantify a match between 
an AI-generated and a human-designed recipe.
During {\it{discovery}},
we use values larger than zero, $\textsf{SDS}>0$, 
to quantify the novelty 
of an AI-generated recipe 
compared to the human-designed recipes in the training set. \\[6.pt]
{\sffamily{\bfseries{Popularity Score to quantify palatability.}}}
Palatability refers to qualities that make a food item desirable to the human palate, such as flavor, aroma, and texture. 
While the human palate displays significant variations across individuals, 
we can still quantify the overall palatability of a food product 
by measuring its {\it{popularity score}} within the population. 
This even extends to patterns in food preparation, 
as evidenced by the popularity of some combinations of ingredients 
compared to others (Fig. \ref{fig1}\textsf{d}). 
Here we propose to use popularity as a proxy for palatability. 
Our AI model 
learns the probability distribution of the human palate
and assigns higher probabilities 
to popular recipes, patterns, and combinations. 
At the recipe level, 
more frequent repetitions 
effectively translate
into a more palatable recipe
associated with a higher popularity score.\\[8.pt]
{\sffamily{\bfseries{\large{Generative AI for burgers}}}} \\[6.pt]
{\sffamily{\bfseries{The Classic Burger.}}}
As a proof of concept, 
we use the generative model to {\it{rediscover}} the \bigmac, 
one of the most widely consumed burgers worldwide \cite{spencer05}, 
served in more than 100 countries \cite{McDonaldsDeliversStrong2011}.  
This global adoption reflects a high degree of palatability 
across diverse populations and 
makes the \bigmac a stringent benchmark 
for evaluating whether the model captures widely shared preferences. 
Because the official recipe is proprietary, 
we approximate it by synthesizing four independent open-source recreations 
into a unified reference recipe 
\cite{taylorstinsonBestHomemadeBig2024, theodorakaloudisBigMac2024, krystlesmithBigMacRecipe2024, rondimpflmaierMakeYourOwn} (Fig.~\ref{fig1}\textsf{b}). 
We then search for this reference recipe 
in randomly generated samples from the model (Fig.~\ref{fig1}\textsf{e}). 
We define rediscovery as a sample 
with a substantial difference score of zero, \textsf{SDS} = 0, 
relative to the reference recipe. 
The training dataset did not contain the reference Big Mac® recipe. \\[6.pt]
{\sffamily{\bfseries{The Delicious Burger.}}}
Next, 
we use our artificial intelligence to {\it{discover}} delicious burgers,
with a pre-defined {\it{novelty score}}.
Specifically,
we adopt the substantial difference score 
to quantify the novelty of an AI-generated sample 
by comparing it to the human-designed recipes in the training set. 
For the \dbone, 
we draw 1 million samples,
filter all samples with 
$\textsf{SDS} \ge 3$,
and select the most repeated sample in the this list
as the most palatable recipe with the highest popularity score
(Fig. \ref{fig2}\textsf{c},\textsf{f}).
For the \dbtwo, 
we perform the same steps,
but now with
$\textsf{SDS} \ge 6$
(Fig. \ref{fig2}\textsf{d},\textsf{g}).
Supplementary File 4 provides details about their preparation.\\[6.pt]
{\sffamily{\bfseries{The Sustainable Burger.}}}
We characterize environmental sustainability 
using life cycle assessment data, 
which estimate the total environmental impact 
of agricultural products 
across production and distribution chains 
based on global producer surveys \cite{hellwegEmergingApproachesChallenges2014}. 
We obtain ingredient-level data 
from a harmonized environmental database 
across $n\,=\,$570 studies \cite{pooreReducingFoodsEnvironmental2018}. 
Since this database does not include mushrooms, 
we supplement it 
with land-use data from the United States Department of Agriculture \cite{Mushrooms2016} and freshwater eutrophication potential, scarcity-weighted water use, and greenhouse gas emissions from European mushroom production \cite{goglioEnvironmentalAssessmentAgaricus2024}. 
We quantify sustainability 
using a single {\it{environmental impact score}} 
that averages normalized land use, aquatic eutrophication potential, scarcity-weighted water use, and greenhouse gas emissions 
across ingredients, weighted by their quantities \cite{clarkEstimatingEnvironmentalImpacts2022}.
For the \sbone, 
a plain mushroom burger with an environmental impact sore of 0.06, 
we draw 1 million samples, 
sort them by their environmental impact score,
and select the most repeated recipe overall
(Fig. \ref{fig3}\textsf{c},\textsf{f}).
For the \sbtwo, 
a beef-mushroom blend with an environmental impact sore of 1.02,
we perform the same steps, 
but now select the most repeated recipe 
that contains both beef and mushroom 
(Fig. \ref{fig3}\textsf{d},\textsf{g}).
Supplementary File~1 reports normalization factors and category assignments. \\[6.pt]
{\sffamily{\bfseries{The Nutritious Burger.}}}
We quantify nutritional quality 
using established nutritional profiling models 
that compare food and nutrient composition against dietary guidelines, 
including the healthy eating index \cite{krebs-smithUpdateHealthyEating2018}, 
nutri-score \cite{juliaFrontofpackNutriScoreLabelling2018}, and 
health star rating \cite{barrett25}. 
Here we use the {\it{healthy eating index}}
developed by the U.S. Department of Agriculture 
to assess alignment with the Dietary Guidelines for Americans
and emphasizes food-group adequacy rather than individual nutrients \cite{20152020DietaryGuidelines2015, herforthGlobalReviewFoodBased2019}. 
%
%
We obtain food-group equivalents 
from the USDA Food Patterns Equivalents Database \cite{bowmanFoodPatternsEquivalents2020} and nutrient composition data from USDA FoodData Central \cite{USDA_FDC_Foundation},
and compute the healthy eating index
by aggregating ingredient-level food-group and nutrient contributions, 
normalized to 500\,kcal servings. 
For the \nutri,
a bean-based formulation
with a healthy eating index of 63.12, 
we draw 1 million samples, 
sort them by their healthy eating index (Fig.~\ref{fig4}\textsf{e}),
and select the most repeated recipe within the top 5\%
(Fig.~\ref{fig4}\textsf{d},\textsf{f}).
%
Supplementary File~1 reports recipes for alternative nutrient scores, and 
Supplementary File~4 provides preparation details.\\[6.pt]
{\sffamily{\bfseries{The Personalized Burger.}}}
We account for inter-individual variation in nutritional requirements using a personalized nutrient profiling model that tailors recipes to age, sex, body composition, and physical activity level \cite{mainardiPersonalizedNutrientProfiling2019}. 
We compute a {\it{personalized nutrition score}} on a 0–100 scale using individual characteristics, including age, sex, body weight, height, and physical activity level. 
We derive nutrient-specific target ranges 
from dietary reference intakes 
and acceptable macronutrient distribution ranges \cite{committeetoreviewthedietaryreferenceintakesforsodiumandpotassiumDietaryReferenceIntakes2019, standingcommitteeforthereviewofthedietaryreferenceintakeframeworkRethinkingAcceptableMacronutrient2024}, 
together with World Health Organization guidelines 
on upper intake limits for sodium, free sugars, and saturated fats \cite{WHOGlobalReport2023, worldhealthorganizationGuidelineSugarsIntake2015, worldhealthorganizationSaturatedFattyAcid2023}. 
We aggregate these targets into a single personalized nutrition score 
for each burger. 
Using this framework,  
we generate personalized burger recipes 
for two representative demographic profiles, 
a 15-year-old, 180\,cm, 80\,kg active male and 
a 70-year-old, 170\,cm, 70\,kg moderately active female 
(Fig.~\ref{fig4}\textsf{g}). 
Supplementary File~1 reports additional personalized recipes. \\[8.pt]
{\sffamily{\bfseries{\large{Burger validation}}}} \\[6.pt]
{\sffamily{\bfseries{Burger preparation.}}}
Our AI-generated recipes specify 
ingredients and quantities only, 
and do not include the processing or cooking steps
needed to prepare the actual burgers. 
We engage an executive chef 
to evaluate the ingredient lists 
and determine appropriate preparation, 
cooking, and assembly procedures 
for each recipe. 
We provide
the final preparation protocols 
to an independent group of chefs, 
who prepare our five AI-generated burgers,
the \dbone, \dbtwo, \sbone, \sbtwo, and \nutri, and
purchase original the \bigmac for comparison,
for the sensory survey. 
Supplementary File~4 includes 
all recipes along with the chef's instructions.\\[6.pt]
{\sffamily{\bfseries{Sensory survey.}}}
We conduct a blind
sensory evaluation with $n$\,=\,101 voluntary participants 
from the general population
at an active restaurant in San Francisco, CA, 
in accordance with Stanford University 
Institutional Review Board guidelines. 
Each parti\-ci\-pant evaluates all six burgers 
on a 7-point Likert scale for 
{\it{overall liking}}, {\it{flavor}}, and {\it{texture}}
\cite{stpierre2024},
and answers check-all-that-apply questions
for {\it{12 flavor-}} and {\it{15 texture}}-related attributes. 
All survey responses are fully anonymized.
Supplementary File~2 provides the full questionnaire.\\[6.pt]
{\sffamily{\bfseries{Demographics.}}}
Of the $n$\,=\,101 participants,
47.5\% are male, 47.5\% female, 3\% non-binary, and 2\% prefer not to say;
22\% are 18-25 years old, 26\% are 26-35, 19\% are 36-45, 18\% are 46-55, and 16\% are older than 55; 
65\% are omnivores and 35\% are flexitarians;
the highest degree of education of 4\% is a high school degree, 24\% college, 50\% batchelor's, 11\% master's, 8\% Ph.D. or higher, and 3\% trade school;
4\% eat burgers every day, 20\% 2-3 times per week, 31\% once a week, 27\% 2-3 times per month, 16\% every 1-2 months, and 3\% 4-5 times per year.  \\[6.pt]
{\sffamily{\bfseries{Power and sample size.}}} 
We select the number of participants, $n$\,=\,101, 
to balance feasibility with statistical power. 
For the Likert-scale ratings, 
this sample size enables detection of small-to-moderate effect sizes 
using two-sided Welch’s t-tests. 
For the binary sensory attributes, 
the sample size provides $>$\,80\% power 
to detect differences of more than 20\% 
between burgers at a significance level of $p$\,$<$\,0.05 using paired comparisons. 
We did not perform an a priori power calculation; 
however, the sample size of $n$\,=\,101
is comparable to or larger than 
those commonly used in consumer surveys of food products.\\[6.pt]
{\sffamily{\bfseries{Statistical analysis.}}} 
We report sensory ratings for overall liking, flavor, and texture 
on a 7-point Likert scale 
as mean\,$\pm$\,standard deviation. 
We use two-sided Welch's t-tests to compare 
the AI-generated burgers against the \bigmac. 
We report binary flavor and texture attributes 
as percentage values and
perform paired comparisons to
assess statistical significance using two-sided binomial tests. 
We do not correct for multiple comparisons, 
as all tests were planned and hypothesis-driven. 
We report statistical significance as $p$\,$<$\,0.05 
(Figs.~\ref{fig2}\textsf{f,g},\ref{fig3}\textsf{f,g},\ref{fig4}\textsf{f}).\\[6.pt]
{\sffamily{\bfseries{Supplementary information.}}}
This manuscript is accompanied by the following supplementary files: 
Supplementary File 1 provides details of data preparation, model architecture, training, and inference, quantifying delicious, sustainable, nutritious, and personalized burgers, and additional burger recipes. 
Supplementary File 2 summarizes the recipes for the AI-generated burgers.
Supplementary File 3 shares the preparation and cooking instructions created by the executive chef.
Supplementary File 4. provides the demographics of the survey participants, the questionnaire for sensory feedback survey, and additional survey outcomes. \\[6.pt]
{\sffamily{\bfseries{Acknowledgements.}}}
We thank 
Executive Chef Justin Schneider 
for his culinary expertise in creating preparation instructions, 
Caroline Cotto from NECTAR at Food System Innovations 
for stimulating discussions,
and
Alice Wistar and Alex Weissman from Palate Insights
for performing the costumer survey.
We acknowledge 
access to the Stanford Marlowe Computing Platform for high performance computing.
This research was supported 
by the Schmidt Science Fellowship, 
in partnership with the Rhodes Trust, to Vahidullah Tac, and
by the Stanford Doerr School of Sustainability Accelerator, 
by the Stanford Bio-X Snack Grant Program,
by the Bezos Earth Fund, 
by the NSF CMMI grant 2320933, and 
by the ERC Advanced Grant 101141626 to Ellen Kuhl. 
\clearpage


\section*{Supplementary material}

# Supplementary File 1

# Generative Artificial Intelligence creates delicious, sustainable, and nutritious burgers

Vahidullah Tac[1*], Christopher Gardner[2] and Ellen Kuhl[1]

[1]Department of Mechanical Engineering, Stanford University, Stanford, USA.
[2]Prevention Research Center, Stanford University School of Medicine, Stanford, USA.

*Corresponding author(s). E-mail(s): vtac@stanford.edu;
Contributing authors: cgardner@stanford.edu; ekuhl@stanford.edu;

## Data preparation

The dataset used in this study is distilled from open-source databases of recipes from `food.com` [1, 2]. The original database at [1] lacks the units of measurements of the ingredients, so we combine it with a complementary database [2]. The combined database consists of 522,517 recipes by 57,178 authors on various categories of foods. Table 1 illustrates a typical recipe from the combined database.

### Identifying burger recipes

There are a few approaches for identifying burger recipes, all with their own caveats. One approach is to simply select recipes that mention the phrase `burger` in the name. However, this fails for some entries such as `hamburger soup`, which is a soup but still contains the term `burger`, or `beef sliders`, which is a burger but does not contain the term `burger`. Another approach is to use a large language model. We use a combination of the two methods to maximize training data points and reduce false positives. For each one of the recipes, a large language model, Mistral 7b [3], is prompted to decide if a recipe is a burger recipe given its name, for each one of the recipes. A recipe is deemed a *burger recipe* if the result of the large language model search is positive positive *and* the name contains the term `burger`.

### Parsing burger ingredients

As shown in Table 1, the ingredients and their quantities and units of measurement are given in plain text. Some ingredient descriptions are non-standard, for example: quantities with multiples, e.g., two 0.5-quart cans; countable items without units, e.g., two eggs; missing quantities, e.g., some salt; ranges, rather than exact values, e.g., one to two tomatoes; and ingredients with descriptors, e.g., fresh oregano. This makes parsing the recipes challenging. To standardize heterogeneous ingredient descriptions across online recipes, we use a large language model to parse raw ingredient text into structured triplets,

$$\{\texttt{ingredient name}, \texttt{quantity}, \texttt{unit}\}.$$

The model normalizes quantities to single numeric values, converts measurements to mass or volume when possible, averages reported ranges, and removes qualitative descriptors such as size or preparation. We

Table 1 **Representative burger recipe.** A typical recipe from the combined `food.com` database contains a recipe ID and name; author ID and name; cooking, preparation, and total times; publication date; description; images; recipe category; keywords; ingredients; recipe instructions; and scores. From a total of 522,517 recipes in the `food.com` database, we identify 3,882 burger recipes from which we generate our training data.

| | |
|---|---|
| Recipe ID | 2734 |
| Name | Ken's Hamburgers |
| Author ID | 1575 |
| Author Name | lindaWWJD |
| Cook Time | 10 min |
| Prep Time | 10 min |
| Total Time | 20 min |
| Date Published | 1999-08-25 05:21:00 |
| Description | Make and share this Ken's Hamburgers recipe from Food.com. |
| Images | https://img.sndimg.com/food/image/upload/w_555,h_416,c_fit,fl_progressive,q_95/v1/img/recipes/27/34/picgsfaLB.jpg, https://img.sndimg.com/food/image/upload/w_555,h_416,c_fit,fl_progressive,q_95/v1/img/recipes/27/34/picaPkIS9.jpg |
| Recipe Category | Meat |
| Keywords | Very Low Carbs, High Protein, Kid Friendly, High In..., Broil/Grill, ¡ 30 Mins, Oven, Easy |
| Ingredients (raw) | • 1/3 lb lean ground beef (per burger)<br>• 1 1/2 teaspoons barbecue sauce<br>• 1/4 teaspoon dehydrated onion, minced |
| Recipe Instructions | • Allow 1/3 lb. lean ground beef per burger.<br>• To each burger, add 1 1/2 teaspoons barbeque sauce and 1/4 teaspoon dehydrated minced onion.<br>• Cook on broiler, turning about every 4 minutes.<br>• Also great on the grill.<br>• Serve with more barbeque sauce, not catsup. |
| Aggregated Rating | 4.0 |
| Review Count | 5 |
| Calories | 273.2 |
| Fat Content | 15.2 |
| ⋮ | ⋮ |

then manually review and correct the parsed outputs to resolve any remaining inconsistencies. The resulting curated dataset contains 3,882 recipes with 3,683 unique ingredients expressed in 171 distinct units of measurement.

## Unifying units of measurement

The units of measurement are processed further for finding commonalities and standardizing the data. Some of the processing steps include: unifying spellings of units of measurement, e.g., g vs gr or teaspoon vs tsp; eliminating recipes that use a unique unit of measurement from consideration; eliminating recipes that use non-standard units of measurement, e.g., can or package; and resolving formatting issues from parsing step. This results in a further simplification to 3,725 recipes utilizing 3,535 unique ingredients measured in 30 unique units of measurement.

## Unifying ingredient names

At this stage, all ingredient names are manually processed to resolve large language model artifacts. Some examples include: simplifying texts, e.g., X becomes Y or X → Y instead of just Y; removing additional explanations from the large language model; removing special characters, e.g.,*, ", . converting plurals to singulars, e.g., leaves to leaf or tomatoes to tomato; unifying multiple spellings of the same ingredient, e.g., chile and chili. The ingredient names are further streamlined by removing unnecessary descriptors: removing

marketing terms and brand names, e.g., from Heinz ketchup to ketchup; grouping all proprietary seasoning mixes and rare seasoning mixes under the overarching term seasoning; combining all beef products, e.g., ground beef, steak, beef chuck, etc. are replaced by beef; combining chicken and pork products; combining variations of the same ingredient name, e.g., adobo and adobo sauce or artichoke and artichoke hearts; combining similar products into one, e.g., baking powder and baking soda; classifying all types of bread into one of three categories, bread, roll, and bun; and correcting misspellings. This resulted in a reduction to only 709 unique ingredients. Next, the recipe list is processed through a first frequency filter by removing all recipes that contain rare ingredients, used less than 10 times, since such low utilization is a barrier for learning about those ingredients. Then, the list of unique ingredients is further standardized including: combining vegetables and fruits into their overarching categories, e.g., kidney beans and pinto beans into beans; combining all vinegar types; combining all liquid oils; combining all fish; combining all types of bread and bun into bun; and eliminated water from the ingredients. Finally, the recipe list is processed through a second frequency filter by removing all recipes that contain rare ingredients, used less than 15 times. At the end of this step, the final curated dataset consists of 2218 recipes and 146 unique ingredients.

### Converting units of measurement into grams

The objective of this step is to replace all units of measurement with grams. For this, the units of measurement undergo the following processing steps: if the quantity is not available, the unit of measurement is removed; all units of volume, e.g., gallons, liter, tablespoon, teaspoon, are converted to milliliter; all units of weight, e.g., kilogram, pound, ounce, are converted into gram; and misattributions of the large language model are manually corrected, e.g., from one cube of beef to one cube of beef bouillon. Next, all ingredient volumes are converted into ingredient weights, from units of milliliter to gram, using density values from food density databases [4–6]. Finally, all remaining measurement units are converted into grams [7]. To do so, ingredient names are paired with unique USDA FDC IDs by searching the database for the ingredient name, and selecting the closest match. This step was facilitated by the `usda_fdc` package in python. The search results were sorted with the following priorities: exact matches; matches with least descriptors, e.g., bouillon cube is preferred over beef bouillon cube when searching for bouillon; matches from `Foundation foods` category; and matches from `SR Legacy foods` category. Table 2 summarizes the final list of ingredients and their FDC IDs.

### Processing ingredient quantities

The automated processing steps with code or large language models detailed so far can result in erroneous entries such as non-numeric entries for quantities, e.g., 1/2 instead of 0.5. These are corrected through manual processing. The recipes are then normalized by calories such that the ingredients of each recipe sum up to 500 kcal using the calorie density of each ingredient from the FDC database. At this final stage, some ingredients in the database can still have missing quantities. We replace these missing values with the average quantity across all recipes that utilize that ingredient. All these steps collectively result in the final `burgers database` where each recipe is represented through

$$\{\texttt{ingredient name}; \texttt{weight in grams}\}.$$

and stored in a 2,216 x 146 array, where the columns correspond to recipes, and the rows correspond to the quantities of the ingredients. Table 2 summarizes all 146 ingredients and their corresponding USDA FDC codes.

Fig. 1 highlights some statistics of the data including the top 10 most used ingredients by frequency and weight. Fig. 2 displays typical relative distributions of weight for the top 10 most used ingredients by frequency in Fig. 1a.

## Modeling

In the current study, a recipe refers to the selection of ingredients and their quantities to be used in a burger. There are a total of 146 unique ingredients (Table 2), and each recipe is a combination of a few of the ingredients in this list. A typical recipe has around ten ingredients and rarely exceeds twenty (Fig. 1, main manuscript). Modeling such data is challenging, as it represents a combination of *discrete* and *continuous* data. Here we propose a sequential two step approach: first, identify the ingredients that will be used in a

**Table 2 List of all 146 ingredients used in this study.** Burger ingredients are down-selected from an initial list of 3,683 by combining ingredient categories to a list of 709, filtered through a first frequency filter, again down-selected by combining ingredient categories, and filtered through a second frequency filter, resulting in the 146 ingredients listed in alphabetic order with their USDA FDC codes [7].

| Ingredient name | FDC ID | Ingredient name | FDC ID | Ingredient name | FDC ID |
| --- | --- | --- | --- | --- | --- |
| allspice | 171315 | curry powder | 170924 | pasta | 168927 |
| almond | 2707486 | dill | 172233 | peanut | 2515376 |
| apple | 2709215 | dough | 172791 | pepper | 170931 |
| arugula | 2709791 | dressing | 173592 | pesto | 171582 |
| avocado | 2709223 | egg | 171287 | pickle | 169379 |
| bacon | 168277 | feta | 173420 | pimento | 168559 |
| barbecue sauce | 174523 | fish | 2706284 | pineapple | 169124 |
| basil | 172232 | flour | 169761 | pork | 2705863 |
| bean | 173744 | garlic | 169230 | portobello | 169255 |
| beef | 2514743 | ginger | 169231 | potato | 170027 |
| beer | 168749 | goat cheese | 2705716 | provolone | 170850 |
| beet | 2685576 | gouda | 171241 | quinoa | 168874 |
| blue cheese | 172175 | green onion | 170005 | ranch sauce | 173592 |
| bouillon | 171562 | greens | 2709792 | relish | 168561 |
| breadcrumb | 1930877 | guacamole | 2709307 | rice | 168879 |
| broth | 172883 | hoisin sauce | 172886 | rosemary | 173473 |
| brown sugar | 2710260 | honey | 169640 | sage | 170935 |
| buffalo | 175299 | horseradish | 173472 | salsa | 174524 |
| bulgur | 2710820 | hot sauce | 171186 | salt | 173468 |
| bun | 2707784 | italian seasoning | 1887966 | sauce | 174524 |
| butter | 173430 | jalapeno | 168576 | sausage | 174584 |
| cabbage | 169975 | ketchup | 168556 | sesame seed | 170150 |
| cajun seasoning | 1887349 | lamb | 2705907 | shallot | 170499 |
| carrot | 2709660 | lemon | 2709168 | sour cream | 2346387 |
| cayenne pepper | 170932 | lentil | 172420 | soy sauce | 174278 |
| celery | 2709778 | lettuce | 2709789 | spinach | 2709614 |
| cheddar cheese | 2705709 | lime | 168155 | steak sauce | 171825 |
| other cheese | 2705764 | liquid smoke | 2680455 | steak seasoning | 2157193 |
| chicken | 2706091 | mango | 2709242 | sugar | 169655 |
| chicken seasoning | 2060346 | marjoram | 170928 | sunflower seed | 170562 |
| chickpea | 173756 | mayonnaise | 168112 | taco seasoning | 172243 |
| chili | 170497 | milk | 2705385 | tahini | 2707587 |
| chili flake | 579086 | mint | 173474 | tarragon | 170937 |
| chip | 2709434 | monterey jack | 2705720 | teriyaki sauce | 171167 |
| chipotle chili | 2009291 | mozzarella | 170845 | thyme | 173470 |
| chive | 2709781 | mushroom | 169251 | tofu | 172476 |
| cilantro | 169997 | mustard | 171043 | tomato | 170457 |
| cinnamon | 171320 | noodle | 2708354 | tortilla | 175036 |
| coriander seed | 170922 | nutmeg | 171326 | turkey | 171505 |
| corn | 2709783 | oat | 2708489 | vinegar | 172241 |
| cornflour | 2710835 | oil | 748278 | walnut | 170187 |
| cottage cheese | 172182 | olive | 2710089 | wasabi | 168583 |
| cracker | 2708184 | onion | 170000 | wheat germ | 168892 |
| cranberry | 2709279 | orange | 169918 | wine | 173185 |
| cream | 2705597 | oregano | 171328 | worcestershire sauce | 171610 |
| cream cheese | 173418 | other seasoning | 171331 | yeast | 167717 |
| cucumber | 2709784 | paprika | 171329 | yogurt | 171284 |
| cumin | 170923 | parmesan | 171247 | zucchini | 168565 |
| curry | 2710178 | parsley | 170416 | | |

recipe using a *mask model*; second, determine the quantities of the chosen ingredients using a *value model*. The first step is a discrete modeling process, while the second step deals with continuous data. We use a multinomial diffusion modeling for the ingredient mask, and a score-based generative model for the ingredient weights.

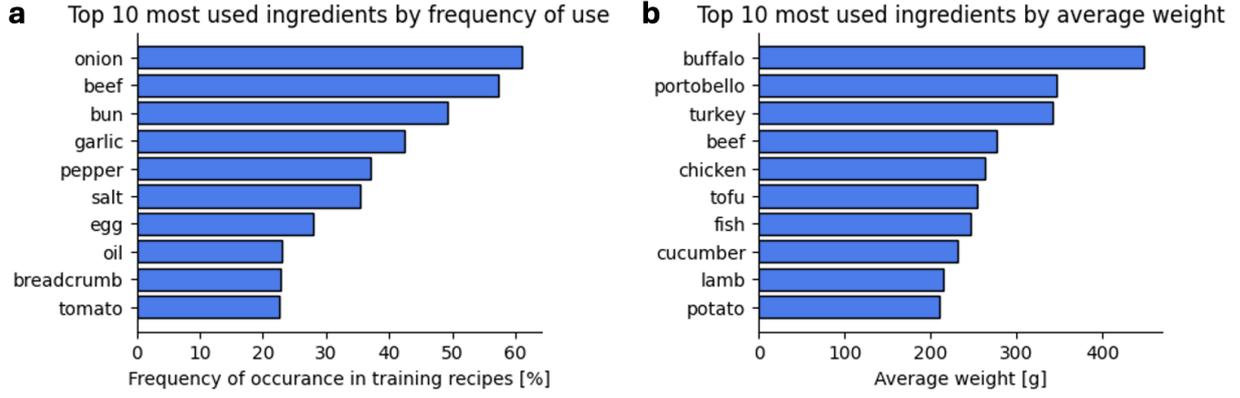

**Fig. 1 Top 10 most used ingredients.** Most used ingredients, out of a total of 146 ingredients, in the 2,216 recipes of our training data, sorted by frequency of use, with onion, beef, and bun being part of more than half of all recipes (a) and by average weight, with buffalo, portobello, and turkey contributing more than 300 g (b).

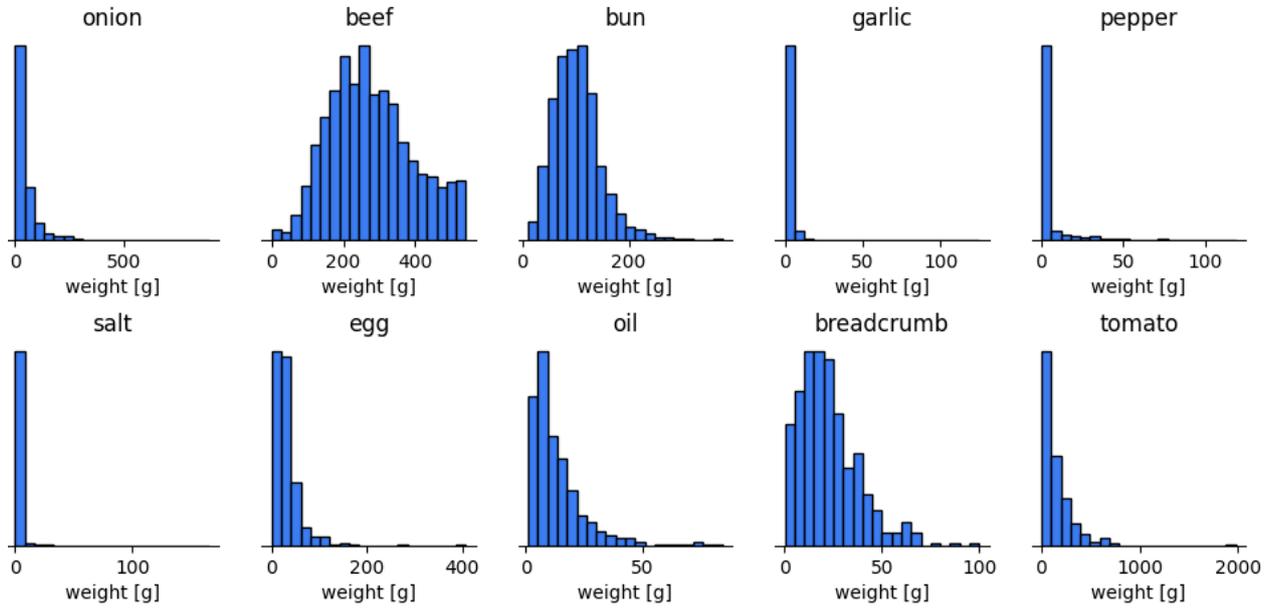

**Fig. 2 Relative distribution of weight of the top 10 ingredients.** Most used ingredients, out of a total of 146 ingredients, in the 2,216 recipes of our training data, sorted by frequency of use (Fig. 1a), reported by their relative distributions.

## Mask model

In multinomial diffusion [8], the diffusion process is defined in terms of categorical distributions that has a $\beta_t$ chance of resampling a category uniformly,

$$q(\mathbf{m}_t|\mathbf{m}_{t-1}) = \mathcal{C}(\mathbf{m}_t|(1-\beta_t)\mathbf{m}_{t-1} + \beta_t/K). \tag{1}$$

As a result, when using a small $\beta_t$, this has a large probability of resampling the same category as $\mathbf{m}_{t-1}$ and a small probability of uniformly selecting a differing category. Since these distributions form a Markov chain, the distribution at any $t$ can be given in terms of the initial data $\mathbf{m}_0$ as

$$q(\mathbf{m}_t|\mathbf{m}_0) = \mathcal{C}(\mathbf{m}_t|\bar{\alpha}_t\mathbf{m}_0 + (1-\bar{\alpha}_t)/K), \tag{2}$$

5

where $\bar{\alpha}_t = \prod_{\tau=1}^{t}(1 - \beta_\tau)$. The posterior distribution has the following form,

$$q(\mathbf{m}_{t-1}|\mathbf{m}_t, \mathbf{m}_0) = \mathcal{C}(\mathbf{x}_{t-1}|\boldsymbol{\theta}_{\text{post}}(\mathbf{m}_t, \mathbf{m}_0)) \quad \text{with} \quad \boldsymbol{\theta}_{\text{post}} = \boldsymbol{\theta}/\sum_{k=1}^{K}\theta, \tag{3}$$

where $\boldsymbol{\theta}_{\text{post}}$ is the normalized vector of the weight vector $\boldsymbol{\theta}$ that combines information from the noisy state $\mathbf{m}_t$ and the origin data $\mathbf{m}_0$,

$$\boldsymbol{\theta} = [(1 - \beta_t)\mathbf{m}_t + \beta_t/K] \odot [\bar{\alpha}_{t-1}\mathbf{m}_0 + (1 - \bar{\alpha}_{t-1})K], \tag{4}$$

where $\odot$ signifies element-wise Hadamard product. Contrary to conventional diffusion models that predict the parameters for predicting $\mathbf{m}_{t-1}$ directly, we instead model the noise which provides an explicit form for the posterior mean and results in a well conditioned loss [9]. Using this approach, we estimate $\mathbf{m}_0$ from $\mathbf{m}_t$ using a neural network $\hat{\mathbf{m}}_0 = \mu(\mathbf{m}_t, t)$. Then, the posterior becomes

$$p(\mathbf{m}_0|\mathbf{m}_1) = \mathcal{C}(\mathbf{m}_0|\hat{\mathbf{m}}_0) \quad \text{and} \quad p(\mathbf{m}_{t-1}|\mathbf{m}_t) = \mathcal{C}(\mathbf{m}_{t-1}|\boldsymbol{\theta}_{\text{post}}(\mathbf{m}_t, \hat{\mathbf{m}}_0)). \tag{5}$$

We use $1,000$ time steps for the diffusion process and adopt a linearly increasing sequence for $\beta_t$ ranging from $\beta(t=0) = 10^{-4}$ to $\beta(t=1000) = 5 \cdot 10^{-3}$.

### Value model

We are interested in modeling the *conditional* distribution $p(\mathbf{w}|\mathbf{m})$, given $N$ samples $\{\mathbf{m}, \mathbf{w}\}^N$, where $\mathbf{m}$ is the recipe mask and $\mathbf{w}$ is the vector of ingredient weights. The conditioning on $\mathbf{m}$ is not explicitly shown in the following derivations for simplicity, but it is implied unless otherwise stated. Score-based generative models are another class of diffusion models where the noising and denoising processes are performed using stochastic differential equations. Here, for the noising process,

$$d\mathbf{w}_t = f(\mathbf{w}_t, t)dt + g(t)dB_t, \tag{6}$$

and for the denoising process,

$$d\mathbf{w}_t = [g^2(t)\nabla_w \log p_t(\mathbf{w}) - f(\mathbf{w}_t, t)]dt + g(t)d\tilde{B}_t, \tag{7}$$

where $B_t$ is $K$-dimensional Brownian motion, and $\tilde{B}_t$ is its time reverse, and $f$ and $g$ are functions that calculate the drift and diffusion processes [10, 11]. While there are a variety of options for the two stochastic differential equations, we use the Ornstein-Uhlenbeck process because of its simplicity [10, 12]. Scaling the stochastic differential equations results in the following set, for the noising process,

$$d\mathbf{w}_t = -\frac{1}{2}\beta_t\mathbf{w}_t dt + \sqrt{\beta_t}d\mathbf{B}_t \tag{8}$$

and for the denoising process,

$$d\mathbf{w}_t = \frac{1}{2}\beta(t)\mathbf{w}_t dt + \beta(t)\nabla\log p_t(\mathbf{w}_t)dt + \sqrt{t\beta(t)}d\mathbf{B}_t, \tag{9}$$

with $\beta(t) = \beta_{\min} + t(\beta_{\max} - \beta_{\min})$. This results in the following simple expression for the conditional distribution,

$$p(\mathbf{w}_t|\mathbf{w}_0) = \mathcal{N}(\mathbf{w}_t|\mu(t)\mathbf{w}_0, \sigma(t)\mathbf{I}) \tag{10}$$

with mean $\mu(t)$ and variance $\sigma(t)$ given as

$$\mu(t) = \exp(-\alpha(t)/2) \quad \text{and} \quad \sigma(t) = 1 - \exp(-\alpha(t)) \quad \text{with} \quad \alpha(t) = \int_0^t \beta(s)ds. \tag{11}$$

The expression $\nabla_w \log p_t(\mathbf{w})$ is known as the *score function*. We approximate the score function by a neural network $s$ parametrized by parameters $\theta$, as $s_{\boldsymbol{\theta}}(\boldsymbol{w}, t) \approx \nabla_\mathbf{w} \log p_t(\boldsymbol{w})$ for the generative model. As a reminder, $\mathbf{x}$ is conditioned on the mask $\mathbf{m}$. When including this explicitly, the neural network approximation takes the form $s_{\boldsymbol{\theta}}(\mathbf{w}, t, \mathbf{m}) \approx \nabla_\mathbf{w} p_t(\mathbf{w}|\mathbf{m})$. To train the neural network, we minimize the following loss

$$L(\boldsymbol{\theta}) = \int_0^T \mathbb{E}_{p_t}\left[||\nabla_\mathbf{w} \log p_t(\mathbf{w}) - s_{\boldsymbol{\theta}}(\mathbf{w}, t)||^2\right], \tag{12}$$

where $p_t(\mathbf{w})$ is approximated using the data [10],

$$\hat{p}_t(\mathbf{w}) = \frac{1}{N} \sum_{i=1}^N p(\mathbf{w}|\mathbf{w}_0^i). \tag{13}$$

We use 1000 steps for the noising and denoising processes and select scaling parameters $\beta_{\min} = 0.001$ and $\beta_{\max} = 3$ [13].

## Training and validation

The *mask model* is constructed using a neural network with an embeddings layer with 1000 embeddings for the time variable, and 3 fully connected layers of 512 neurons each for predicting the logits of the mask model. The logits are then used with the softmax function to produce the raw probabilities. We use minibatching with $n = 1000$ for training and train with a learning rate of $5 \cdot 10^{-4}$ using the Adam optimization algorithm for 100,000 epochs.

The *value model* is constructed using a feed-forward neural network with 4 hidden layers of 256 neurons each. The vectors for $\mathbf{x}, t$ and $\mathbf{m}$ are concatenated to form the input for $s_{\boldsymbol{\theta}}$. We use a batch size of 400 and a learning rate of $1 \cdot 10^{-3}$ with the Adam optimization algorithm and train for 20,000 epochs. The training data is split 80% for training and 20% for validation. Training is conducted on Nvidia H100 and L40S GPUs.

## Rediscovering the Big Mac®

The original Big Mac® weighs 210 g and has 563 kcal. Since its recipe is proprietary, we approximate it using open-source recreations [14–17]. We identify the common ingredients at the intersection of the four masks,

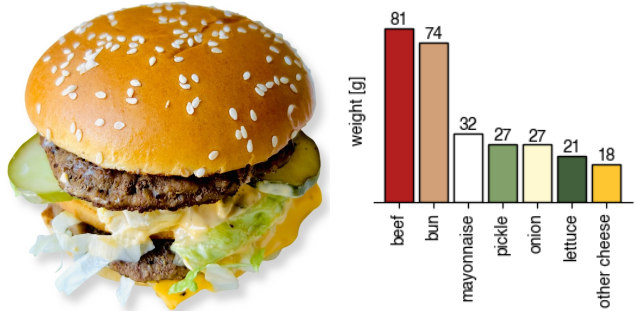

**Fig. 3 Classic Big Mac® as a benchmark for rediscovery.** Since the original recipe of the Big Mac® (left) is proprietary, we use online recipes to estimate its ingredients and weights as 81 g ground beef, 74 g bun, 32 g mayonnaise, 27 g pickle, 27 g onion, 21 g lettuce, and 18 g cheese (right).

and average their quantities after normalizing by calories. The resulting Big Mac® recipe for a 500 kcal serving consists of seven ingredients with the following weights: 81 g ground beef, 74 g bun, 32 g mayonnaise, 27 g pickle, 27 g onion, 21 g lettuce, and 18 g cheese (Fig. 1**b** main manuscript, Fig. 3).

7

## Substantial Difference Score

To quantify the proximity between recipes and grouping similar recipes we introduce the Substantial Difference Score (SDS). We define the substantial difference between two recipes $r_1$ and $r_2$ as the sum of the binary distance $d_i$ over all ingredients in the database,

$$\text{SDS}(r_1, r_2) = \sum_{i=1}^{n_{\text{ing}}} d_i(r_1, r_2), \quad \text{with} \quad d_i(r_1, r_2) = \begin{cases} 1 & \text{if } r_{1i} + r_{2i} \neq 0 \text{ and } r_{1i} \cdot r_{2i} = 0 \\ 1 & \text{if } \max(r_{1i}, r_{2i})/\min(r_{1i}, r_{2i}) \geq 2 \\ 0 & \text{otherwise.} \end{cases} \quad (14)$$

The substantial difference score is zero, $\text{SDS} = 0$, if two recipes share the same ingredients and the ratios of their quantities never exceed two. Two recipes $r_1$ and $r_2$ with a substantial difference score of zero are called repeats or a match. During rediscovery, we use the substantial difference score of zero, $\text{SDS} = 0$, to identify a match between an AI-generated recipe and a human-designed recipe. During discovery, we use values larger than zero, $\text{SDS} > 0$, to quantify the novelty of an AI-generated recipe compared to the human-designed recipes in the training set. The Delicious Burger 1 of the main manuscript, is the most repeated recipe with $\text{SDS} = 3$ (Fig. 2c,f, main manuscript). It contains 151 g ground beef, 79 g tomato, 75 g dill pickle, 47 g yellow onion, 30 g brioche bun, 19 g mayonnaise, 12 g american cheese, 7 g lettuce, 7 g ketchup, and 1 g salt. The Delicious Burger 2 of the main manuscript, is the most repeated recipe with $\text{SDS} = 6$ (Fig. 2d,g, main manuscript). It contains 62.0 g ground beef, 35.5 g bun, 28.0 g sliced tomato, 12.5 g remoulade sauce, 8.0 g Gruyère cheese, 3.5 g yellow onion, 2.0 g Worcestershire sauce, 1.0 g brown sugar, 0.5 g salt, 0.2 g garlic, and 0.2 g fresh thyme leaves.

## Environmental Impact Score

Life cycle assessments estimate the total environmental impact of agricultural products across their production and distribution chains by surveying producers worldwide [18]. Here we quantify sustainability using the environmental database [19], a meta-analysis that harmonizes data from $n = 570$ studies. As this database does not include mushrooms, we supplement it with the averaged land-use data for cultivated mushrooms from the United States Department of Agriculture [20], and freshwater eutrophication potential, scarcity-weighted water use, and greenhouse gas emissions from three European countries [21]. To enable optimization and comparison, we condense environmental impacts into a single Environmental Impact Score (EIS) [22]. We compute EIS by normalizing and averaging impacts across four categories: land use (LU), aquatic eutrophication potential (EU), scarcity-weighted water use (WU), and greenhouse gas emissions (GHG) as

$$\text{EIS} = \frac{1}{4}\left[\frac{\text{LU}}{\text{LU}_{\max}} + \frac{\text{EU}}{\text{EU}_{\max}} + \frac{\text{WU}}{\text{WU}_{\max}} + \frac{\text{GHG}}{\text{GHG}_{\max}}\right] \quad (15)$$

where the values of $\text{LU}_{\max}$, $\text{EU}_{\max}$, $\text{WU}_{\max}$, and $\text{GHG}_{\max}$ were directly obtained from [22] as 62.2 m$^2$, 20.2 kg CO2 equivalents, 174 g PO$_4$ equivalents, and 43,600 L. From this information, we calculate the environmental impact score for each of the ingredients in this study (Fig. 4). We quantify the environmental impact of a recipe by summing the ingredient-level environmental impact score contributions, weighted by ingredient quantities. We manually match the ingredients from in our model (Table 2) to one of the categories in the environmental database (Table 3) [19]. Most ingredients in this study are whole agricultural products; however, a small number of processed ingredients, such as condiments, are composed of multiple components. Given the lack of consistent environmental data and their negligible contribution by mass, we exclude condiments, bouillons, broths, yeast, and vinegar from the environmental impact calculations.

Environmentally sustainable burgers were identified by finding recipes that are repeated often in a given range of EIS scores. For example, we can generate burger recipes with $0.1 \leq \text{EIS} < 0.2$ and $0.2 \leq \text{EIS} < 0.3$ (Fig. 5a,b). The Sustainable Burger 1 of the main manuscript, a plain mushroom burger with an environmental impact sore of 0.06, is the most repeated recipe with $\text{EIS} < 0.1$. (Fig. 3c,f, main manuscript). It contains 226 g portobello mushroom 78 g mayonnaise, 46 g brioche bun, 21 g arugula, 9 g spicy brown mustard, 8 g extra virgin olive oil, 3 g rosemary, 2 g fresh garlic, and 1 g kosher salt. The Sustainable Burger 2 of the main manuscript, a beef-mushroom blend with an environmental impact sore of 0.51, is the most repeated recipe that contains both beef and mushroom (Fig. 3d,g, main manuscript). It contains 70 g ground beef, 42 g

8

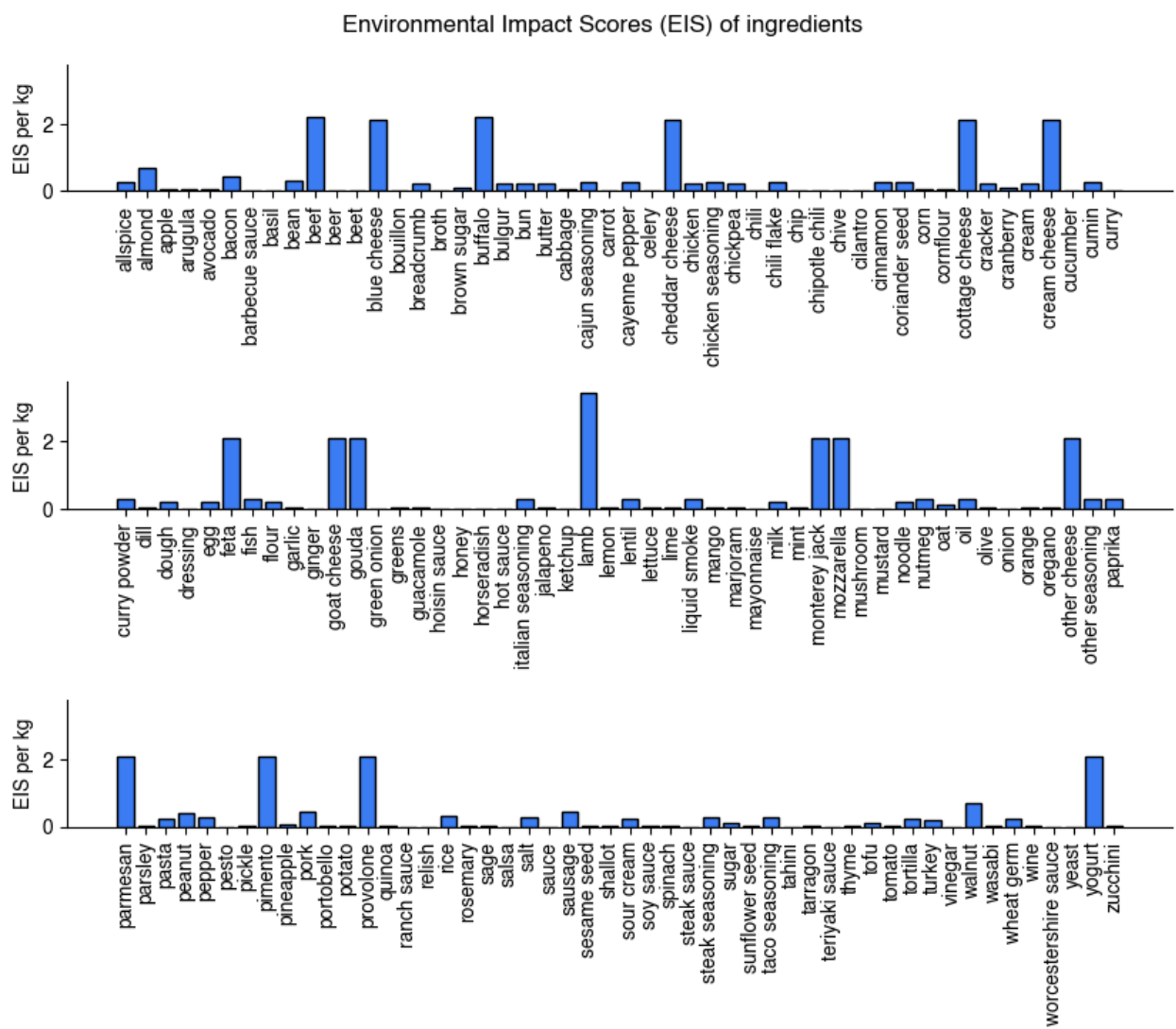

**Fig. 4 Environmental impact scores of ingredients.** Reported environmental impact scores measure the collective impact per 1 kg for each of the 146 ingredients in our database across four categories: land use, aquatic eutrophication potential, scarcity-weighted water use, and greenhouse gas emissions.

brioche bun, 39 g cheddar cheese, 19 g bacon, 17 g ketchup, 14 g crimini mushroom, 9 g yellow onion, 1 g black pepper, and 1 g kosher salt.

## Healthy Eating Index

Nutritional profiling models quantify the nutritional quality of foods and diets by comparing nutrient and food-group composition against dietary guidelines. Common nutritional profiling models include the Healthy Eating Index (HEI) [23], the Nutri-Score [24], and the Health Star Rating [25]. The Healthy Eating Index was developed by the U.S. Department of Agriculture to assess alignment with the Dietary Guidelines for Americans [26] on a 0–100 scale, with higher scores indicating closer adherence. It primarily emphasizes food-group adequacy, such as fruits, vegetables, whole grains, and dairy, rather than individual nutrients, consistent with global trends toward food-based dietary guidelines [27]. In contrast, other nutritional profiling models explicitly emphasize nutrients of concern, including protein, fiber, potassium, calcium, iron, and vitamin D [28–30], and, in some cases, bioactive compounds such as flavonoids [31]. Here we use the healthy eating index as the primary metric to evaluate the nutritional quality of our generated burger recipes. We

**Table 3 Mapping of our 146 burger ingredients onto environmental database LCA categories.** To calculate the environmental impact score, we map each the 146 ingredients from in our model (Table 2) to one of the LCA categories in the environmental database.

| LCA Category | Burger Ingredients |
| --- | --- |
| Wheat & Rye (Bread) | breadcrumb, bulgur, bun, cracker, dough, flour, noodle, pasta, tortilla, wheat germ |
| Maize (Meal) | corn, cornflour |
| Barley (Beer) | beer |
| Oatmeal | oat |
| Rice | rice |
| Potatoes | chip, potato |
| Sugar | brown sugar, sugar |
| Other Pulses | bean, lentil |
| Peas | chickpea |
| Nuts | almond, walnut |
| Groundnuts | peanut |
| Soymilk | soy sauce |
| Tofu | tofu |
| Oil | oil |
| Tomatoes | tomato |
| Onions & Leeks | green onion, chive, onion, shallot |
| Root Vegetables | beet, carrot, ginger, horseradish, wasabi |
| Brassicas | arugula, cabbage, lettuce |
| Other Vegetables | basil, celery, chili, chipotle chili, cilantro, cucumber, dill, garlic, greens, jalapeno, marjoram, mint, oregano, parsley, pickle, quinoa, rosemary, sage, sesame seed, spinach, sunflower seed, tarragon, thyme, zucchini |
| Citrus Fruit | lemon, lime, orange |
| Apples | apple |
| Berries & Grapes | cranberry |
| Wine | wine |
| Other Fruit | avocado, mango, guacamole, olive, pineapple |
| Beef | beef |
| Lamb & Mutton | lamb |
| Pig Meat | bacon, pork, sausage |
| Poultry Meat | chicken, turkey |
| Milk | milk |
| Cheese | blue cheese, cheddar cheese, cottage cheese, cream cheese, feta, goat cheese, gouda, monterey jack, mozzarella, other cheese, parmesan, pimento, provolone, yogurt |
| Eggs | egg |
| Fish (farmed) | fish |
| Sweeteners & Honey | honey |
| Stimulants & Spices Misc. | allspice, cajun seasoning, cayenne pepper, chicken seasoning, chili flake, cinnamon, coriander seed, cumin, curry powder, italian seasoning, liquid smoke, nutmeg, other seasoning, paprika, pepper, salt, steak seasoning, taco seasoning |
| Buffalo | buffalo |
| Butter, Cream & Ghee | butter, cream, sour cream |
| Mushrooms | portobello, mushroom |
| Condiments & Others | barbecue sauce, curry, dressing, hoisin sauce, hot sauce, ketchup, mayonnaise, mustard, pesto, ranch sauce, relish, salsa, sauce, steak sauce, tahini, teriyaki sauce, worcestershire sauce, bouillon, broth, yeast, vinegar |

sample a large number of recipes from the generative model and identify representative nutritious burgers by selecting the most frequently occurring recipe among those in the 90th and 95th percentiles of the healthy eating index distribution. Selecting highly repeated samples ensures that the resulting recipes correspond to regions of high probability under the generative model and are therefore likely to be palatable and popular. The Nutritious Burger of the main manuscript, a bean-based burger with a healthy eating index of 63.12, is the most repeated recipe in the 95th percentile of the healthy eating index (Fig. 4d, main manuscript). It contains 71 g canned kidney beans, 50 g brioche bun, 15 g rolled oats, 10 g egg, 4 g corn flour, 4 g extra virgin olive oil, 2 g jalapeño, 2 g cilantro, 2 g white onion, 0.8 g kosher salt 0.2 g cumin, 0.2 g black pepper, and

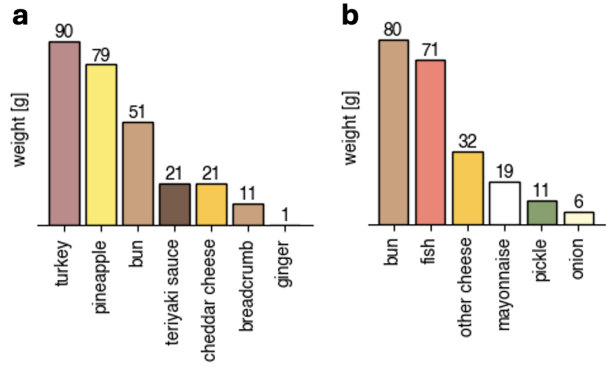

**Fig. 5 Alternative sustainable burgers.** Most repeated recipes with an environmental impact score of $0.1 \leq \mathsf{EIS} < 0.2$ (a) and $0.2 \leq \mathsf{EIS} < 0.3$ (b), ranking between the Sustainable Burger 1 with $\mathsf{EIS} = 0.03$ and Sustainable Burger 2 with $\mathsf{EIS} = 0.51$ of the main manuscript.

0.05 g Mexican oregano. For comparison, is the most repeated recipe in the 90th percentile of the healthy eating index is a bean-corn blend with different ingredients and nutritional profile (Fig. 6a,b).

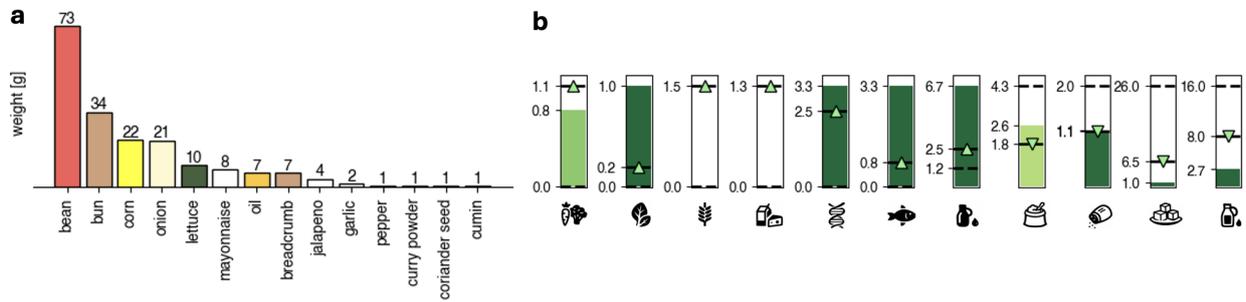

**Fig. 6 Alternative nutritious burger.** Most repeated recipe in the 90th percentile of the healthy eating index (HEI), a bean-corn blend with ingredients (a) and nutritional profile (b).

In addition, we use the balanced hybrid nutrient density score (bHNDS) [28, 32], a combined food-group- and nutrient-based nutritional profiling model, to generate alternative nutritious burger formulations. For comparison, we generated the 95th and 90th percentile nutritious burger recipes by balanced hybrid nutrient density score (Fig. 7a,b).

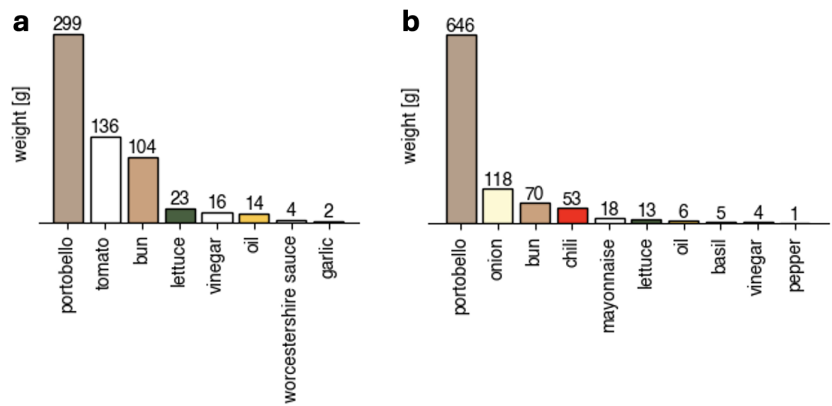

**Fig. 7 Alternative nutritious burgers.** Most repeated recipes in the 95th percentile (a), and 90th percentile (b) by balanced hybrid nutrient density score (bHNDS).

11

## Personalized nutrition

Nutritional requirements vary substantially across individuals as a function of age, sex, body composition, physical activity, and physiological status. For example, dietary reference intakes for iron in adult women aged 19–50 years are more than twice those for men of the same age group [33]. To account for such heterogeneity, we adopt a *personalized nutrient profiling model* [34] that enables the generation of recipes tailored to individual nutritional requirements. We compute a personalized nutrition score on a 0–100 scale based on individual characteristics, including age, sex, body weight, height, and physical activity level. We derive recommended intake ranges from dietary reference intakes and acceptable macronutrient distribution ranges [33, 35], together with World Health Organization guidelines on upper intake limits for sodium, free sugars, and saturated fats [36–38]. These recommendations jointly define nutrient-specific target ranges that are aggregated into a single *personalized nutrition score* for each burger. Using this framework, we generate personalized burger recipes for four representative demographic profiles: a 30-year-old sedentary male of 80 kg and 180 cm, a 30-year-old sedentary female of 60 kg and 160 cm, a very active 15-year-old male of 55 kg and 160 cm, and a moderately active 70-year-old female of 70 kg and 170 cm (Fig. 8).

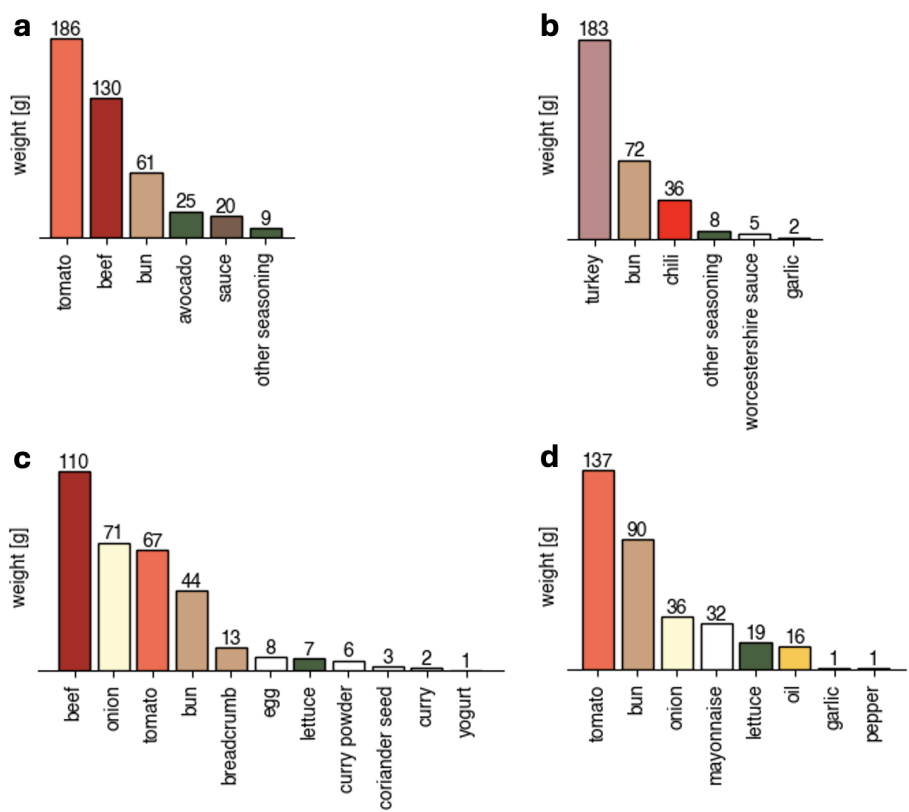

**Fig. 8 Personalized nutritious burgers.** Personalized recipes for a sedentary 30 year old male weighing 80 kg with a 180 cm height (a), a sedentary 30 year old female weighing 60 kg with a 160 cm height (b), a very active 15 year old male weighing 55 kg with a 160 cm height (c), and a moderately active 70 year old female weighing 70 kg with a 170 cm height (d).

# Supplementary File 2

# Generative Artificial Intelligence creates delicious, sustainable, and nutritious burgers

Vahidullah Tac[1*], Christopher Gardner[2] and Ellen Kuhl[1]

[1]Department of Mechanical Engineering, Stanford University, Stanford, USA.
[2]Prevention Research Center, Stanford University School of Medicine, Stanford, USA.

*Corresponding author(s). E-mail(s): vtac@stanford.edu;
Contributing authors: cgardner@stanford.edu; ekuhl@stanford.edu;

## Sensory evaluation

We enrolled $n = 101$ voluntary participants from the general population for a blind sensory evaluation of all six burgers at an active restaurant in San Francisco, CA. The study was performed in accordance with Stanford University Institutional Review Board guidelines. We asked the participants to answer seven question about their background, and then, for each burger, rank overall liking, flavor, and texture on a 7-point Likert scale, and check-all-that-apply for 12 flavor attributes and 15 texture attributes.

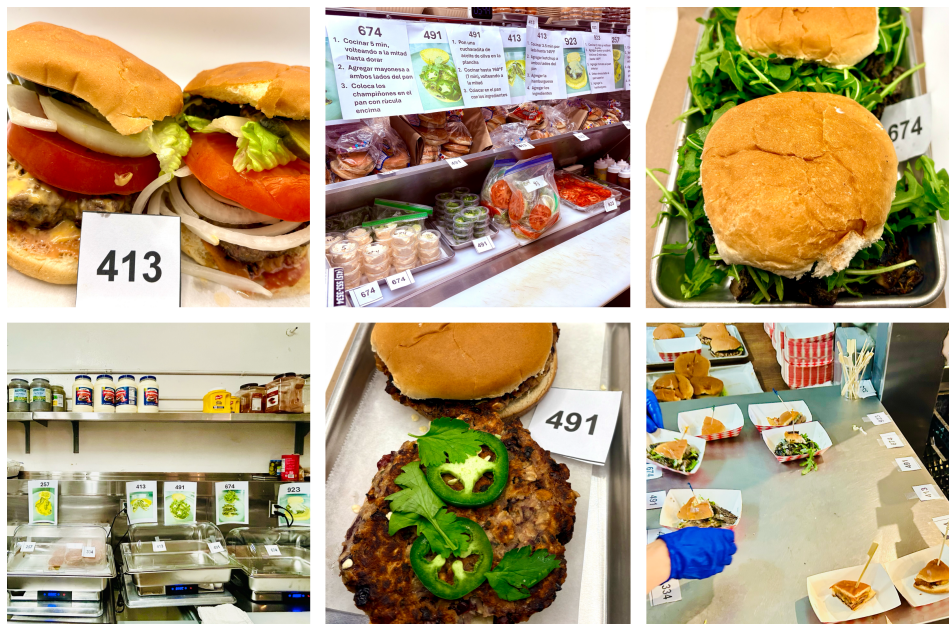

**Fig. 1 Sensory evaluation.** Voluntary participants from the general population enrolled in a blind sensory evaluation of all six burgers in an active restaurant. Snapshots during the preparation of all six burgers and of Sustainable Burger 1 labeled as 413, Delicious Burger 1 labeled as 674, and Nutritious Burger labeled as 491, prior to sampling.

# Participant information

## What is your age?

Select one

- Less than 18
- 18-25
- 26-35
- 36-45
- 46-55
- Greater than 55

## Which gender do you identify as?

Select one

- Male
- Female
- Non-binary
- Prefer not to say

## What is the highest degree or level of education you have completed?

Select one

- Some high school
- High school
- Some college
- Bachelor's degree
- Master's degree
- Ph.D. or higher
- Trade school

## How often do you eat burgers of any kind?

Select one

- Everyday
- 2-3 times a week
- Once a week
- 2-3 times a month
- Once every 1-2 months
- 4-5 times a year
- 2-3 times a year
- Never or rarely

## What is your dietary preference?

Select one

- Vegan
- Vegetarian
- Prescatarian
- Flexitarian (i.e. eats meat occasionally)
- Omnivore (i.e. eats meat regularly)

## Where do you shop regularly for groceries?

Select all that apply

- Natural grocers (e.g. Whole Foods)
- Discount grocers (e.g. Trader Joe's, Lidl, Aldi)
- Conventional grocers (e.g. Stop & Shop, Safeway)
- Big box retailers (e.g. Target, Walmart)
- Grocery delivery services (e.g. Instacart, Amazon fresh)
- Meal kits (e.g. Blue Apron)
- Bulk grocers (e.g. Costco, Sam's Club)
- Small neighborhood grocers
- Directly from the brand/manufacturer's website

## What matters most to you when deciding what to eat?

Select all that apply

- Health
- Price
- Taste
- Convenience
- Familiarity (what I'm used to)
- Environment
- Animal welfare

# Sensory feedback

## How would you rate your overall liking of Burger X?

Select one

- Like very much
- Like
- Like somewhat
- Neither like nor dislike
- Dislike somewhat
- Dislike
- Dislike very much

## How would you rate the flavor of Burger X?

Select one

- Like very much
- Like
- Like somewhat
- Neither like nor dislike
- Dislike somewhat
- Dislike
- Dislike very much

## Please check all the words or phrases that describe the flavor of Burger X.

Select all that apply

- Meaty
- Weird Aftertaste
- Off-flavor
- Bland
- Earthy / Soil
- Smoky
- Strong
- Sweet

- Salty
- Fatty
- Savory
- Good aftertaste

**How would you rate the texture of Burger X?**

Select one

- Like very much
- Like
- Like somewhat
- Neither like nor dislike
- Dislike somewhat
- Dislike
- Dislike very much

**Please check all the words or phrases that describe the texture of Burger X.**

Select all that apply

- Chewy
- Crispy / Crunchy
- Crumbly / Grainy
- Firm / Hard
- Soft / Mushy
- Holds together
- Moist
- Dry
- Tough
- Fatty
- Fibrous / Stringy
- Brittle
- Gummy
- Springy
- Sticky

## Population demographics

The $n=101$ study participants represent a balanced cross section of the general population (Fig. 2). Of the $n=101$ participants, 47.5% were male, 47.5% female, 3% non-binary, and 2% prefer not to say. 22% were 18-25 years old, 26% are 26-35, 19% are 36-45, 18% are 46-55, and 16% are older than 55. 65% were omnivores and 35% were flexitarians. Their highest degree of education was 4% high school degree, 24% college, 50% batchelor's, 11% master's, 8% Ph.D. or higher, and 3% trade school. 4% eat burgers every day, 20% 2-3 times per week, 31% once a week, 27% 2-3 times per month, 16% every 1-2 months, and 3% 4-5 times per year. For grocery purchasing habits, 37% preferred natural grocers, 55% discount grocers, 58% conventional grocers, 12% big box retailers, 12% grocery delivery services, 30% bulk grocers, 26% small neighborhood grocers, and 2% manufacturer's websites. For purchasing priorities, 50% listed health, 61% price, 76% taste, 25% convenience, 16% familiarity, 20% environment, and 18% animal welfare.

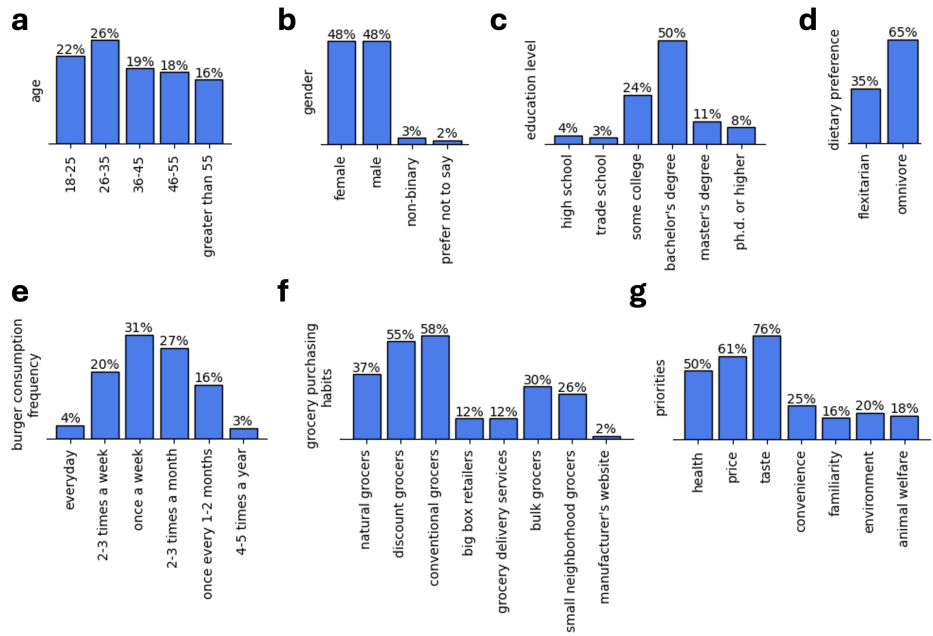

**Fig. 2 Population demographics.** Overview of the demographics and food preferences of the survey respondents. Participant age, gender, education, dietary preference, burger consumption frequency, grocery purchasing habits, and priorities from the responses to the first seven survey questions.

# Supplementary File 3

# Generative Artificial Intelligence creates delicious, sustainable, and nutritious burgers

Christopher Gardner[2] and Ellen Kuhl[1]

[1]Department of Mechanical Engineering, Stanford University, Stanford, USA.
[2]Prevention Research Center, Stanford University School of Medicine, Stanford, USA.

Contributing authors: cgardner@stanford.edu; ekuhl@stanford.edu;

## Guidelines for Recipe Development

This document was handed to an Executive Chef. It contains the AI-generated ingredient lists and their corresponding quantities, and describes how the lists should be interpreted.

## Recipes

This study includes 5 burgers. The ingredients and their quantities for the 5 burgers are given below. Guidelines for the Chefs follow in the next section.

### Delicious Burger 1

- beef 151 g
- bun 30 g
- ketchup 7 g
- lettuce 7 g
- mayonnaise 19 g
- onion 47 g
- other cheese 12 g
- pickle 75 g
- tomato 79 g

### Delicious Burger 2

- beef 124 g
- brown sugar 2 g
- bun 71 g
- garlic 0.4 g
- onion 7 g
- other cheese 16 g
- other sauce 25 g
- thyme 0.3 g
- tomato 56 g
- worcestershire sauce 4 g

### Sustainable Burger 1

- arugula 21 g
- bun 46 g
- garlic 2 g
- mayonnaise 78 g
- mustard 9 g
- oil 8 g
- portobello 226 g
- rosemary 3 g

### Sustainable Burger 2

- bacon 19 g
- beef 70 g
- bun 42 g
- cheddar cheese 39 g
- ketchup 17 g
- mushroom 14 g
- onion 9 g
- black pepper 1 g

### Nutritious Burger

- bean 71 g
- bun 50 g
- cilantro 2 g
- cornflour 4 g
- cumin 0.2 g
- egg 10 g
- jalapeno 2 g
- oat 15 g
- oil 4 g
- onion 2 g
- oregano 0.05 g
- black pepper 0.2 g

## Guidelines

1. The lists in the previous section show the ingredient amounts that should be used in each burger. The chefs should develop recipes with these ingredients and amounts, without adding or removing any ingredients.
2. The ingredient lists only show the ingredients, but all processing steps are to be decided by the chefs.
3. All the ingredients are listed in singular words, e.g., oat instead of oats, onion instead of onions, etc. This is simply for convenience.
4. The chefs should decide which parts of the ingredient lists should be used in the patty, and which parts should be layered on top. This could include dividing some ingredients into two or more groups. For example, if the recipe mentions onion, the chef could decide to mince some of it and add to the patty, and slice some for layering on top of the patty. In such cases, we ask the chefs to record the weights of the parts.
5. The amounts given in each recipe are adjusted such that each ingredient list adds up to 500 calories. The chefs are free to adjust the quantities such that the ingredients make up one burger instead, as long as the ratio of the ingredients remains unchanged. For example, if the chef believes that the ingredients listed for a burger would make 2 burgers, they are free to divide the amounts of the ingredients by 2 as long as this is done for each ingredient.
6. The weights are generally meant to be dry and uncooked weights. If the chef believes that the weight listed for an ingredient is too high/low for it to be the dry/uncooked weight, e.g., beans, please contact us know for further clarification.
7. If the amount for a herb or spice is extremely low, the chef can reinterpret it as a really small amount, e.g., 0.05 g of oregano can be interpreted as a couple of leaves of oregano.

8. salt: the ingredient lists do not explicitly list salt. The chefs should adjust salt to taste, and record the amount they used.
9. All of the listed ingredients are meant to be used in preparing the burgers, but not all of them have to stay on the burger. Examples:
    - Vegetables and fruits have peels and stems that can be removed after weighing.
    - If the recipe lists oil, and if the chef decides that the oil is best used for frying or sauteing, the oil may remain in the pan after sauteing
    - Some herbs like rosemary may be used to add aroma to the rest of the ingredients and disposed afterwards.
10. Some of the recipes may result in overflowing burgers, this is acceptable but the chefs are still kindly asked to try their best in preparing the burgers in a way that is presentable, such that the burger has a structural integrity.
11. The ingredient lists do not mention specific details like variety of the ingredients. The chefs are free to make their own decisions about this. Examples:
    - Beef may be interpreted as intact meat or gound beef, depending on what the chef believes is the best choice for that particular recipe.
    - Vegetables can be any variety that the chef prefers, e.g., red onions, white onions, etc., but not green onions, which are a different product.
12. Other cheese includes any cheese that is not one of these: blue cheese, cheddar cheese, cottage cheese, cream cheese, feta, goat cheese, gouda, monterey, monterey jack, mozzarella, parmesan, provolone. We ask the chef to select a cheese that they think best fits the rest of the ingredients.
13. Other sauce includes any sauce that is not one of these: barbecue sauce, buffalo sauce, cream, curry sauce, hoisin sauce, honey, hot pepper sauces, ketchup, mustard, pesto, ranch, relish, salsa, soy sauce, steak sauce, teriyaki sauce, vinegar, wasabi, Worcestershire sauce. We ask the chef to select a sauce that they think best fits the rest of the ingredients.
14. In Sustainable Burger 2, we ask that the chef combine the meat and mushrooms together to make the patty. This way the burger is more aligned with commercially available beef-mushroom blends.
15. We kindly ask the chefs to put their best creativity and artistic skills to work to turn these ingredient lists to the best burgers possible, and wholeheartedly thank them for their help in this study!

# Supplementary File 4

# Generative Artificial Intelligence creates delicious, sustainable, and nutritious burgers

Vahidullah Tac[1*], Christopher Gardner[2] and Ellen Kuhl[1]

[1]Department of Mechanical Engineering, Stanford University, Stanford, USA.
[2]Prevention Research Center, Stanford University School of Medicine, Stanford, USA.

*Corresponding author(s). E-mail(s): vtac@stanford.edu;
Contributing authors: cgardner@stanford.edu; ekuhl@stanford.edu;

## Recipes with ingredients and preparation instructions

This document was prepared by an Executive Chef. It contains the AI-generated ingredient lists, their quantities, and the preparation steps and photos of the assembled burgers as designed by the chef. For comparison, we also include the ingredient list of the Big Mac®, although we purchased it for the survey to provide the most authentic experience.

## Big Mac®

### List of ingredients

- 81 g ground beef
- 74 g bun
- 32 g mayonnaise
- 27 g pickle
- 27 g onion
- 21 g lettuce
- 18 g cheese

### Purchased product

Figure 1 shows the Big Mac®as a benchmark comparison in the sensory survey. The Big Mac®was not prepared by the chefs but purchased to provide the most authentic sensory experience.

## Delicious Burger 1

### List of ingredients

- 151 g ground beef (80/20)
- 30 g brioche bun, trimmed significantly to make weight
- 7 g ketchup
- 7 g lettuce, little gem

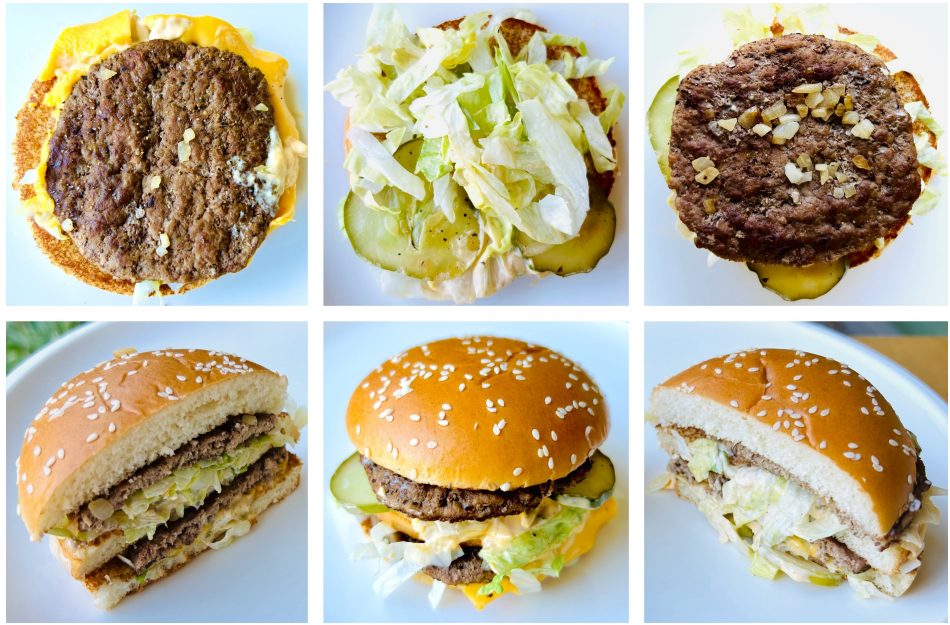

**Fig. 1 Big Mac®.** The Big Mac® is classic beef burger with seven main ingredients, ground beef, mayonnaise, pickle, onion, lettuce, and cheese, served in a bun.

- 19 g mayo
- 47 g yellow onion, sliced rings
- 12 g American cheese, slice
- 75 g dill pickle, chips
- 79 g tomato, sliced
- 1 g kosher salt

## Preparation instructions

Figure 2 shows the Delicious Burger 1 prepared according to the following instructions: Using a ring mold slightly larger than the bun, shape the mixture into a patty, pressing it into an even layer inside the mold. Season both sides of the patty with salt. Preheat a nonstick skillet over medium heat. Lightly toast the bun in the dry pan until golden brown, then remove and set aside. Add the patty to the hot pan and cook for 3.5 to 4 minutes per side, adding the cheese during the last minute to allow it to melt. In a small bowl, mix the mayonnaise and ketchup until combined, then spread the sauce evenly on both sides of the bun. Assemble the burger by stacking the lettuce, tomato, pickles, and onions carefully on top of the patty.

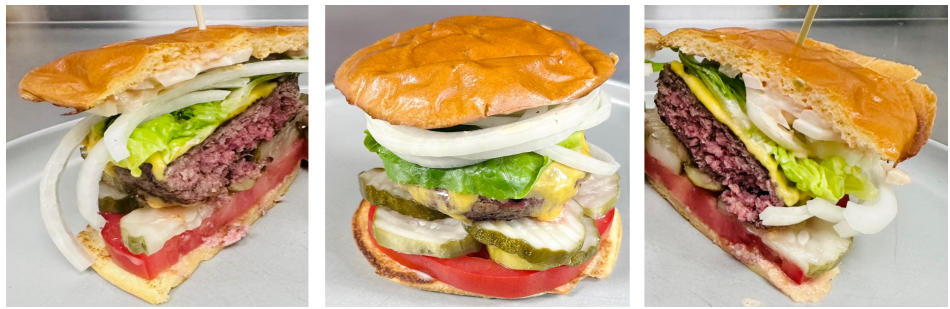

**Fig. 2 Delicious Burger 1.** The Delicious Burger 1 is a classical beef burger optimized for deliciousness with nine ingredients, beef, tomato, pickle, onion, mayo, cheese, lettuce and ketchup, served in a bun.

# Delicious Burger 2

## List of ingredients
- 62 g ground beef (80/20)
- 1 g brown sugar
- 35.5 g bun
- 0.2 garlic clove, microplaned
- 3.5 g yellow onion, shaved
- 8 g Gruyère cheese, sliced
- 12.5 g rémoulade sauce
- 0.15 g fresh thyme leaves, minced
- 28 g sliced tomato
- 2 g Worcestershire sauce
- 0.5 g kosher salt

## Preparation instructions

Figure 3 shows the Delicious Burger 2 prepared according to the following instructions: In a bowl, thoroughly mix the beef with the garlic, thyme, brown sugar, and Worcestershire sauce. Using a ring mold slightly bigger than the bun, shape the mixture into a patty pressing it into an even layer inside the mold. Preheat a nonstick pan over medium-high heat. Lightly toast the bun in the dry pan until golden brown, then remove and set aside. Season the burger patty with salt. Arrange the sliced onions on top of the patty, then place it in the hot pan onion-side down. Cook for 2 minutes, then flip the patty, add the cheese, and cook for another 2 minutes. Cover the pan with a lid to help the cheese melt. To assemble, layer the sliced tomato on the bottom bun, followed by the burger patty. Spread the remoulade sauce on the top bun, then close the burger

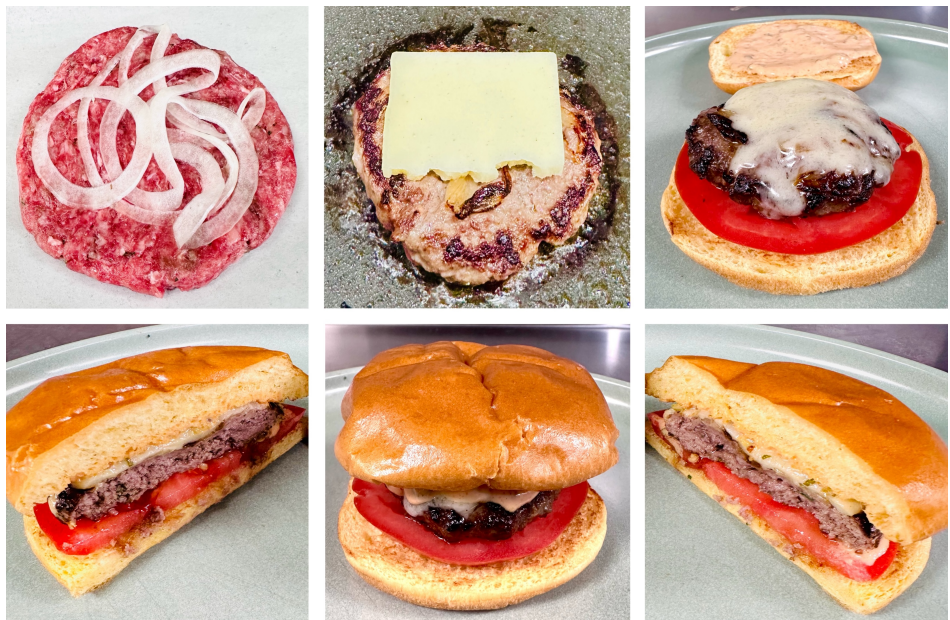

**Fig. 3 Delicious Burger 2.** The Delicious Burger 2 is an innovative beef burger optimized for deliciousness with ten ingredients, beef, tomato, remoulade, Gruyère cheese, onion, Worcestershire sauce, brown sugar, garlic, and thyme leaves, served in a bun.

# Sustainable Burger 1

## List of ingredients
- 21 g arugula

- 46 g brioche bun (trimmed to make weight)
- 2 g fresh garlic, microplaned
- 78 g mayonnaise, divided (Best Foods)
- 9 g spicy brown mustard
- 8 g extra virgin olive oil
- 226 g portobello mushroom (stems and gills removed)
- 3 g rosemary, minced
- 0.5 g kosher salt

### Preparation instructions

Figure 4 shows the Sustainable Burger 1 prepared according to the following instructions: Slice portobello into 0.5 inch thick slices across the cap. In a bowl, mix together the garlic, mustard, rosemary, salt, and half of the mayonnaise. Add the mushrooms and fold gently to coat. Let marinate for 20 minutes. Preheat a nonstick pan over medium-high heat. Lightly toast the bun in the dry pan until golden, then remove and set aside. Add the olive oil to the same pan. Once hot, add the marinated mushroom slices and caramelize for about 2.5 minutes per side, until deeply browned. Transfer the mushrooms to a paper towel to drain any excess oil. Spread the remaining mayonnaise on both sides of the bun. Layer the mushrooms on the bottom half and top with fresh arugula.

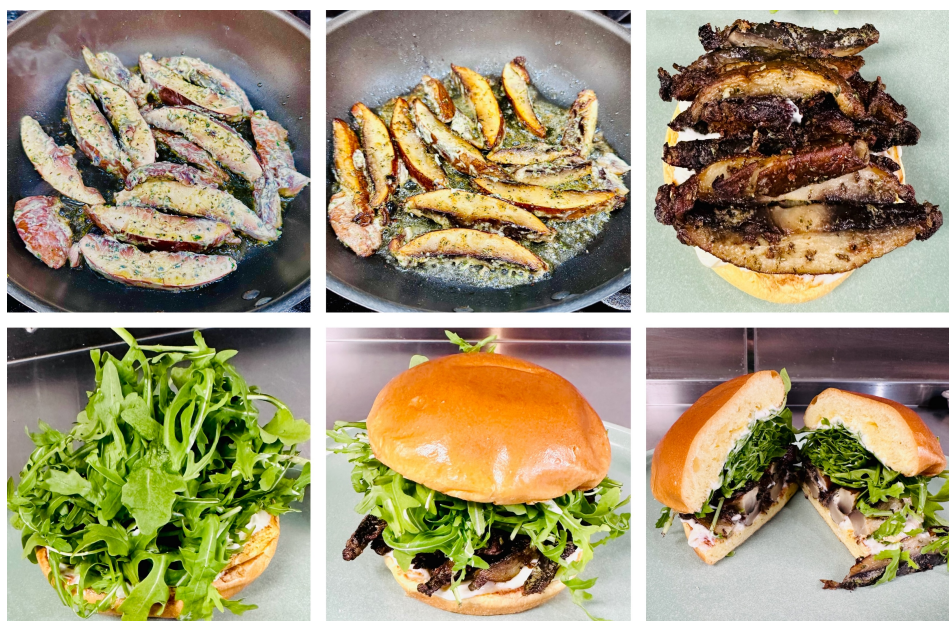

**Fig. 4  Sustainable Burger 1.** The Sustainable Burger 1 is a mushroom burger optimized for sustainability with eight ingredients, portobello mushroom, mayonnaise, arugula, mustard, oil, rosemary, and garlic, served in a bun.

## Sustainable Burger 2

### List of ingredients

- 19 g bacon, cut into two pieces
- 70 g ground beef (80/20)
- 42 g brioche bun, slightly trimmed to make weight
- 39 g cheddar cheese, sliced
- 17 g ketchup
- 14 g crimini mushroom, finely minced
- 9 g yellow onion, shaved
- 1 g black pepper, ground

- 0.8 g kosher salt

## Preparation instructions

Figure 5 shows the Sustainable Burger 2 prepared according to the following instructions: Combine the minced mushrooms and ground beef, folding them together until evenly mixed. Shape the mixture into a ball, then roll it in salt and pepper to coat. Preheat a nonstick skillet over medium heat. Lightly toast the bun in the dry pan until golden, then remove and set aside. Reduce the heat to medium-low, add the bacon, and cook until the fat renders and the bacon is crisp. Transfer the bacon to a paper towel–lined plate to drain, leaving the rendered fat in the pan. Increase the heat to medium-high. Place the sliced onions in the center of the skillet, then set the seasoned meat ball on top. Cover with a piece of parchment or wax paper, and firmly smash the patty with a spatula until it's slightly wider than the bun. Cook for 1.5 minutes, then flip, top with cheese, and cook for another 1.5 minutes, or until the patty is cooked through and the cheese has melted. Spread ketchup on both sides of the toasted bun. Place the burger patty on the bottom half, layer with bacon, and top with the other half of the bun.

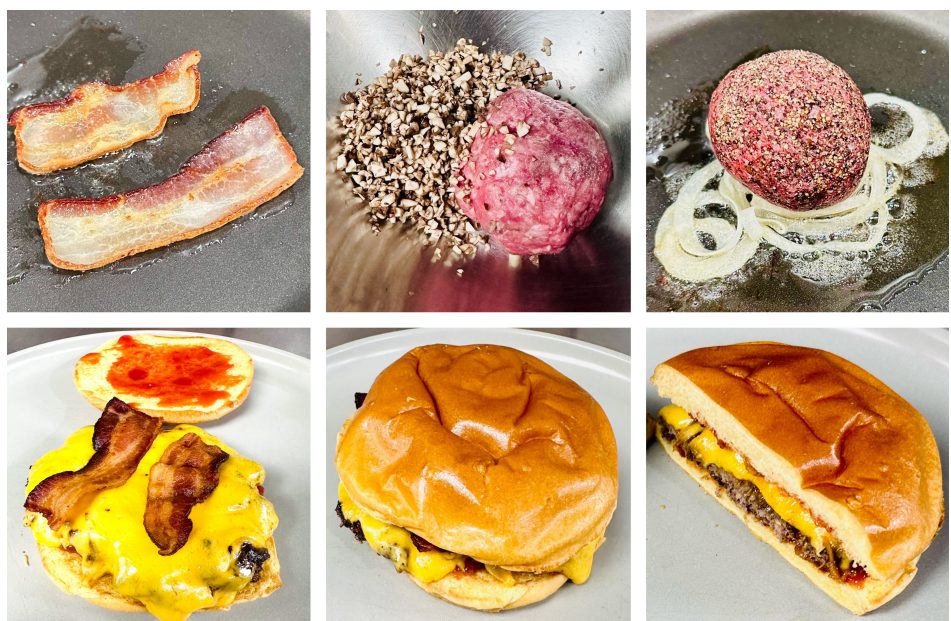

**Fig. 5 Sustainable Burger 2.** The Sustainable Burger 2 is a mushroom-beef blend with eight ingredients, ground beef, cheddar cheese, bacon, ketchup, crimini mushroom, onion, and pepper, served in a bun.

# Nutritious Burger

## List of ingredients

- 71 g canned kidney beans (drained and rinsed)
- 50 g brioche bun, lightly toasted
- 2 g cilantro (leaves and tender stems)
- 4 g corn flour
- 0.2 g cumin, ground
- 10 g egg (whisk one whole egg and measure out 10 g)
- 2 g jalapeño, thinly sliced rounds
- 15 g rolled oats
- 4 g extra virgin olive oil
- 2 g white onion, minced
- 0.05 g Mexican oregano, dried
- 0.2 g black pepper, ground

- 0.8 g kosher salt, divided between the patty mix and seasoning the formed patty before cooking

## Preparation instructions

Figure 6 shows the Nutritious Burger prepared according to the following instructions: Lightly mash the beans with a fork, leaving some chunks for texture. Stir in the onion, cumin, half of the salt, black pepper, and oregano. Add the oats and corn flour, mixing until evenly combined. Finally, add the egg and mix thoroughly to form a cohesive mixture. Using a ring mold the same size as your bun, shape the mixture into a patty on a piece of parchment paper. Refrigerate for 30 minutes to allow the starches to hydrate and the patty to firm up before cooking. Preheat a nonstick pan over medium-high heat and lightly toast the bun in the dry pan. Once the bun is toasted, add the olive oil to the pan. Season the chilled patty with the remaining salt and cook for about 3.5 minutes per side, until golden and heated through. Assemble the burger by placing the cooked patty on the toasted bun, then topping with the remaining cilantro and jalapeño slices.

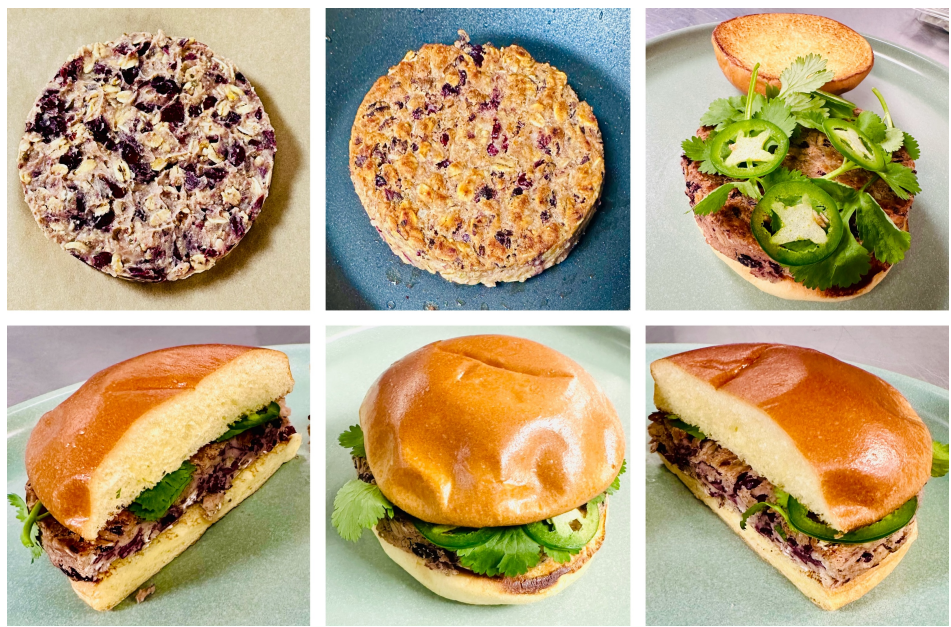

**Fig. 6 Nutritious Burger.** The Nutritious Burger is a bean-based burger optimized for nutrition with twelve ingredients, kidney beans, rolled oats, egg, cornflour, oil, cilantro, jalapeño, onion, cumin, pepper, and oregano, served in a bun.


\end{document}